\documentclass[twocolumn,aps,pra]{revtex4-1}
\usepackage{amsmath,amssymb}
\usepackage[pdftex]{graphicx} 
\usepackage{epstopdf} 
\DeclareGraphicsExtensions{.pdf,.eps,.png,.jpg,.jpeg,.mps} 
\usepackage{epsfig} 
\usepackage{epic}
\usepackage{hyperref} 
\usepackage{color}
\usepackage{fancyhdr}
\usepackage{bm}
\usepackage{lipsum}

\graphicspath{{gfx/}}

\def\op{\hat{o}}

\def\a{\hat{a}}

\def\bd{b^\dagger}
\def\c{\hat{c}}
\def\cd{\hat{c}^\dagger}
\def\n{\hat{n}}
\def\N{\hat{N}}
\def\h{\hat{h}}
\def\H{\hat{H}}

\begin{document} 

\title{Non-Hermitian Dynamics in the Quantum Zeno Limit} 
\author{W. Kozlowski}
\author{S. F. Caballero-Benitez}
\author{I. B. Mekhov}
\affiliation{Department of Physics, Clarendon Laboratory, University
  of Oxford, Parks Road, Oxford OX1 3PU, United Kingdom}

\begin{abstract}
  We show that weak measurement leads to unconventional quantum Zeno
  dynamics with Raman-like transitions via virtual states outside the
  Zeno subspace. We extend this concept into the realm of
  non-Hermitian dynamics by showing that the stochastic competition
  between measurement and a system's own dynamics can be described by
  a non-Hermitian Hamiltonian. We obtain a solution for ultracold
  bosons in a lattice and show that a dark state of tunnelling is
  achieved as a steady state in which the observable's fluctuations
  are zero and tunnelling is suppressed by destructive matter-wave
  interference.
\end{abstract}

\maketitle

\section{Introduction}

Frequent measurements can slow the evolution of a quantum system
leading to the quantum Zeno effect \cite{Zeno1977, Facchi2008} which
has been successfully observed in a variety of systems
\cite{Itano1990, Nagels1997, Kwiat1999, Balzer2000, Streed2006,
  Hosten2006, Bernu2008}. One can also devise measurements with
multi-dimensional projections which lead to quantum Zeno dynamics
where unitary evolution is uninhibited within this degenerate
subspace, i.e.~the Zeno subspace \cite{Facchi2008, Raimond2010,
  Raimond2012, Signoles2014}. In this paper we go beyond conventional
quantum Zeno dynamics. By considering the case of measurement near,
but not in, the projective limit the system is still confined to a
Zeno subspace, but intermediate transitions are allowed via virtual
Raman-like processes. We show that this can be approximated by a
non-Hermitian Hamiltonian thus extending the notion of quantum Zeno
dynamics into the realm of non-Hermitian quantum mechanics joining the
two paradigms.

Non-Hermitian systems exhibit a variety of rich behaviour, such as
localisation \cite{Hatano1996, Refael2006}, $\mathcal{PT}$ symmetry
\cite{Bender1998,Giorgi2010,Zhang2013}, spatial order
\cite{Otterbach2014}, or novel phase transitions \cite{Lee2014a,
  Lee2014b}. Recent experimental results further motivate the study of
these novel phenomena \cite{Dembowski2001, Choi2010, Ruter2010,
  Barontini2013, Gao2015}. Non-Hermitian Hamiltonians commonly arise
in systems with decay or loss \cite{Dhar2015, Stannigel2014}, limited
by possibilities of controlling dissipation. Additionally, the
non-unitary time evolution is subject to discontinuous jumps applied
whenever decay events are detected requiring the postselection of
trajectories \cite{Otterbach2014, Lee2014a, Lee2014b}. Here we
consider systems where the non-Hermitian term arises from measurement
and uncover a novel general mechanism that is independent of the
nature of the original Hamiltonian. It does not rely on losses or
postselection of exotic trajectories, thus conceptually simplifying
experimental realizability of such intriguing effects. Furthermore, we
show that the physics can be much more complicated and lead to
dynamics beyond the conventional Hermitian and quantum Zeno dynamics
paradigms.

As an example, we demonstrate the effects of non-Hermitian evolution
by investigating a gas of ultracold bosons in a lattice inside an
optical cavity which are subjects of intensive interdisciplinary
research \cite{Bloch2007,Lewenstein2007,Ritsch2013,Mekhov2012}. This
field is extremely flexible when it comes to engineering measurement
\cite{Ruostekoski1997, Mekhov2007NP, Chen2009, Mekhov2009LP,
  Mekhov2010LP, Mekhov2011LP, Mekhov2012, Mekhov2013LP, Pedersen2014,
  Elliott2015, Kozlowski2015, Mazzucchi2015, Atom2015, Wade2015,
  Ashida2015, Mazzucchi2016, Caballero2016}. Furthermore, it is a
realistic experimental proposal for our theoretical model given the
rapid progress in merging quantum optics with ultracold gases
\cite{Baumann2010, Wolke2012, Schmidt2014, Miyake2011, Weitenberg2011}
including the most recent realizations of an optical lattice in a
high-Q cavity \cite{Landig2015a, Klinder2015}. We go beyond quantum
nondemolition approaches where the measurement backaction or many-body
dynamics were neglected \cite{Mekhov2007, Mekhov2007PRA,
  PolzikNatPh2008, Roscilde2009, Mekhov2009, Mekhov2009PRA,
  DeChiara2011, Hauke2013, Rogers2014}. We will show that,
counter-intuitively, non-Hermitian dynamics causes two competing
processes, tunnelling and measurement, to cooperate to form a dark
state of the atomic dynamics with zero fluctuations in the observed
quantity without the need for an effective cavity potential which is
typically considered in self-organisation \cite{Nagy2008,
  Niedenzu2013, Bakhtiari2015, Ivanov2015, Schutz2015, Winterauer2015,
  Piazza2015}.

\section{Theoretical Model - Measurement Induced Non-Hermitian
  Dynamics}

\subsection{Suppression of Coherences in the Density Matrix by Strong
  Measurment}

We consider a state described by the density matrix $\hat{\rho}$ whose
isolated behaviour is described by the Hamiltonian $\H_0$ and when
measured the jump operator $\c$ is applied to the state at each
detection \cite{MeasurementControl}. The master equation describing
its time evolution when we ignore the measurement outcomes is given by
\begin{equation}
  \dot{\hat{\rho}} = -i [ \H_0 , \hat{\rho} ] + \c \hat{\rho} \cd - \frac{1}{2}(
  \cd\c \hat{\rho} + \hat{\rho} \cd\c ).
\end{equation}
We also define $\c = \lambda \op$ and $\H_0 = K \h$. The exact
definition of $\lambda$ and $K$ is not so important as long as these
coefficients can be considered to be some measure of the relative size
of these operators. They would have to be determined on a case-by-case
basis, because the operators $\c$ and $\H_0$ may be unbounded. If
these operators are bounded, one can simply define them such that
$||\op|| \sim \mathcal{O}(1)$ and $||\h|| \sim \mathcal{O}(1)$. If
they are unbounded, one possible approach would be to identify the
relevant subspace in whose dynamics we are interested in and scale the
operators such that the eigenvalues of $\op$ and $\h$ in this subspace
are $\sim \mathcal{O}(1)$.

We will use projectors $P_m$ which have no effect on states within a
degenerate subspace of $\c$ ($\op$) with eigenvalue $c_m$ ($o_m$), but
annihilate everything else. For convenience we will also use the
following definition $\hat{\rho}_{mn} = P_m \hat{\rho} P_n$ (these are
submatrices of the density matrix, which in general are not single
matrix elements). Therefore, we can write the master equation that
describes this open system as a set of equations
\begin{align}
\label{eq:master}
  \dot{\hat{\rho}}_{mn} = & -i K P_m \left[ \h \sum_r \hat{\rho}_{rn} - \sum_r \hat{\rho}_{mr} \h \right] P_n \nonumber \\ & + \lambda^2 \left[ o_m o_n^* - \frac{1}{2} \left( |o_m|^2 + |o_n|^2 \right) \right] \hat{\rho}_{mn},
\end{align}
where the first term describes coherent evolution whereas the second
term causes dissipation. 

Firstly, note that for the density submatrices for which $m = n$,
$\hat{\rho}_{mm}$, the dissipative term vanishes and they are thus
decoherence free subspaces and will form the Zeno
subspaces. Interestingly, any state that consists only of these
decoherence free subspaces, i.e.~
$\hat{\rho} = \sum_m \hat{\rho}_{mm}$, and that commutes with the
Hamiltonian, $[\hat{\rho}, \hat{H}_0] = 0$, will be a steady state.
This can be seen by substituting this ansatz into
Eq. \eqref{eq:master} which yields $\dot{\hat{\rho}}_{mn} = 0$ for all
$m$ and $n$. These states can be prepared dissipatively using known
techniques \cite{Diehl2008}, but it is not required that the state be
a dark state of the dissipative operator as is usually the case.

Secondly, we consider a large detection rate, $\lambda^2 \gg K$, for
which the coherences, i.e.~ the density submatrices $\hat{\rho}_{mn}$
for which $m \ne n$, will be heavily suppressed by
dissipation. Therefore, we can adiabatically eliminate these
cross-terms by setting $\dot{\hat{\rho}}_{mn} = 0$, to get
\begin{equation}
\label{eq:intermediate}
\hat{\rho}_{mn} = \frac{K}{\lambda^2} \frac{i P_m \left[ \h \sum_r \hat{\rho}_{rn} - \sum_r \hat{\rho}_{mr} \h \right] P_n } {o_m o_n^* - \frac{1}{2} \left( |o_m|^2 + |o_n|^2 \right)}
\end{equation}
which tells us that they are of order $K/\lambda^2 \ll 1$. One can
easily recover the projective Zeno limit by considering
$\lambda \rightarrow \infty$ when all the subspaces completely
decouple. However, it is crucial that we only consider,
$\lambda^2 \gg K$, but not infinite. If the subspaces do not decouple
completely, then transitions within a single subspace can occur via
other subspaces in a manner similar to Raman transitions. In Raman
transitions population is transferred between two states via a third,
virtual, state that remains empty throughout the process. By avoiding
the infinitely projective Zeno limit we open the option for such
processes to happen in our system where transitions within a single
Zeno subspace occur via a second, different, Zeno subspace even though
the occupation of the intermediate states will remian negligible at
all times.

In general, a density matrix can have all of its $m = n$ submatrices,
$\hat{\rho}_{mm}$, be non-zero and non-negligible even when the
coherences are small. However, for a pure state this would not be
possible. To understand this, consider the state $| \Psi \rangle$ and
take it to span exactly two distinct subspaces $P_a$ and $P_b$
($a \ne b$). This wavefunction can also be written as
$| \Psi \rangle = P_a | \Psi \rangle + P_b | \Psi \rangle$. The
corresponding density matrix is thus given by
\begin{align}
  \hat{\rho}_\Psi = & P_a | \Psi \rangle \langle \Psi | P_a + P_a | \Psi
                      \rangle \langle \Psi | P_b \nonumber
  \\ & + P_b | \Psi \rangle \langle \Psi | P_a +
       P_b | \Psi \rangle \langle \Psi | P_b.
\end{align}
If the wavefunction has significant components in both subspaces then
in general the density matrix will not have negligible coherences,
$\hat{\rho}_{ab} = P_a | \Psi \rangle \langle \Psi | P_b$. Therefore,
a density matrix with small cross-terms between different Zeno
subspaces can only be composed of pure states that each lie
predominantly within a single subspace.

Therefore, in order for the coherences to be of order $K/\lambda^2$ we
would require the wavefunction components to satisfy
$P_a | \Psi \rangle \approx \mathcal{O}(1)$ and
$P_b | \Psi \rangle \approx \mathcal{O}(K/\lambda^2)$. This in turn
implies that the population of the states outside of the dominant
subspace (and thus the submatrix $\hat{\rho}_{bb}$) will be of order
$\langle \Psi | P_b^2 | \Psi \rangle \approx
\mathcal{O}(K^2/\lambda^4)$. Therefore, these pure states cannot exist
in a meaningful coherent superposition in this limit. This means that
a density matrix that spans multiple Zeno subspaces has only classical
uncertainty about which subspace is currently occupied as opposed to
the uncertainty due to a quantum superposition.
 
\subsection{Quantum Measurement vs. Dissipation}

This is where quantum measurement deviates from dissipation. If we
have access to a measurement record we can infer which Zeno subspace
is occupied, because we know that only one of them can be occupied at
any time. The time needed to determine the correct state is (see
Appendix \ref{appendix})
\begin{equation}
  t \gg \frac{1}{\lambda^2} \frac{|o_n|^2}{\left( |o_m|^2 - |o_n|^2
    \right)^2 } \quad \forall m,n \quad m \ne n,
\end{equation}
which is faster than the system's internal dynamics as long as the
eigenvalues are distinguishable enough. Thanks to measurement we can
make another approximation. If we observe a number of detections
consistent with the subspace $P_m = P_0$ we can set
$\hat{\rho}_{mn} \approx 0$ for all cases when both $m \ne 0$ and
$n \ne 0$ leaving our density matrix in the form
\begin{equation}
  \label{eq:approxrho}
  \hat{\rho} = \hat{\rho}_{00} + \sum_{r\ne0} (\hat{\rho}_{0r} +
  \hat{\rho}_{r0}).
\end{equation}
We can do this, because the other states are inconsistent with the
measurement record. We know from the previous section that the system
must lie predominantly in only one of the Zeno subspaces and when that
is the case, $\hat{\rho}_{0r} \approx \mathcal{O}(K/\lambda^2)$ and
for $m \ne 0$ and $n \ne 0$ we have
$\hat{\rho}_{mn} \approx \mathcal{O}(K^2/\lambda^4)$. Therefore, this
amounts to keeping first order terms in $K/\lambda^2$ in our
approximation.

This is a crucial step as all $\hat{\rho}_{mm}$ matrices are
decoherence free subspaces and thus they can all coexist in a mixed
state decreasing the purity of the system without
measurement. Physically, this means we exclude trajectories in which
the Zeno subspace has changed (measurement isn't fully projective). By
substituting Eq. \eqref{eq:intermediate} into Eq. \eqref{eq:master} we
see that this happens at a rate of $K^2 / \lambda^2$. However, since
the two measurement outcomes cannot coexist any transition between
them happens in discrete transitions (which we know about from the
change in the detection rate as each Zeno subspace will correspond to
a different rate) and not as continuous coherent evolution. Therefore,
we can postselect in a manner similar to Refs. \cite{Otterbach2014,
  Lee2014a, Lee2014b}, but our requirements are significantly more
relaxed - we do not require a specific single trajectory, only that it
remains within a Zeno subspace. Furthermore, upon reaching a steady
state, these transitions become impossible as the coherences
vanish. This approximation is analogous to optical Raman transitions
where the population of the excited state is neglected. Here, we can
make a similar approximation and neglect all but one Zeno subspace
thanks to the additional knowledge we gain from knowing the
measurement outcomes.

\subsection{The Non-Hermitian Hamiltonian}

Rewriting the master equation using $\c = c_0 + \delta \c$, where
$c_0$ is the eigenvalue corresponding to the eigenspace defined by the
projector $P_0$ which we used to obtain the density matrix in
Eq. \eqref{eq:approxrho}, we get
\begin{equation}
\label{eq:finalrho}
\dot{\hat{\rho}} = -i \left( \H_\mathrm{eff} \hat{\rho} - \hat{\rho} \H_\mathrm{eff}^\dagger \right) + \delta \c \hat{\rho} \delta \cd,
\end{equation}
\begin{equation}
\label{eq:Ham}
\H_\mathrm{eff} = \H_0 + i \left( c_0^*\c - \frac{|c_0|^2}{2} - \frac{\cd\c}{2} \right).
\end{equation}
The first term in Eq. \eqref{eq:finalrho} describes coherent evolution
due to the non-Hermitian Hamiltonian $\H_\mathrm{eff}$ and the second
term is decoherence due to our ignorance of measurement outcomes. When
we substitute our approximation of the density matrix
$\hat{\rho} = \hat{\rho}_{00} + \sum_{r\ne0} (\hat{\rho}_{0r} +
\hat{\rho}_{r0})$ into Eq. \eqref{eq:finalrho}, the last term
vanishes, $\delta \c \hat{\rho} \delta \cd = 0$. This happens, because
$\delta \c P_0 \hat{\rho} = \hat{\rho} P_0 \delta \c^\dagger = 0$. The
projector annihilates all states except for those with eigenvalue
$c_0$ and so the operator $\delta \c = \c - c_0$ will always evaluate
to $\delta \c = c_0 - c_0 = 0$. Recall that we defined
$\hat{\rho}_{mn} = P_m \hat{\rho} P_n$ which means that every term in
our approximate density matrix contains the projector $P_0$. However,
it is important to note that this argument does not apply to other
second order terms in the master equation, because some terms only
have the projector $P_0$ applied from one side,
e.g.~$\hat{\rho}_{0m}$. The term $\delta \c \hat{\rho} \delta \cd$
applies the fluctuation operator from both sides so it does not matter
in this case, but it becomes relevant for terms such as
$\delta \cd \delta \c \hat{\rho}$.

It is important to note that this term does not automatically vanish,
but when the explicit form of our approximate density matrix is
inserted, it is in fact zero. Therefore, we can omit this term using
the information we gained from measurement, but keep other second
order terms, such as $\delta \cd \delta \c \rho$ in the Hamiltonian
which are the origin of other second-order dynamics. This could not be
the case in a dissipative system.

Ultimately we find that a system under continuous measurement for
which $\lambda^2 \gg K$ in the Zeno subspace $P_0$ is described by the
deterministic non-Hermitian Hamiltonian $\H_\mathrm{eff}$ in
Eq. \eqref{eq:Ham} and thus obeys the following Schr\"{o}dinger
equation
\begin{equation}
 i \frac{\mathrm{d} | \Psi \rangle}{\mathrm{d}t} = \left[\H_0 + i \left(
      c_0^*\c - \frac{|c_0|^2}{2} - \frac{\cd\c}{2} \right) \right] |
  \Psi \rangle.
\end{equation}
Of the three terms in the parentheses the first two represent the
effects of quantum jumps due to detections (which one can think of as
`reference frame' shifts between different degenerate eigenspaces) and
the last term is the non-Hermitian decay due to information gain from
no detections. It is important to emphasize that even though we
obtained a deterministic equation, we have not neglected the
stochastic nature of the detection events. The detection trajectory
seen in an experiment will have fluctuations around the mean
determined by the Zeno subspace, but there simply are many possible
measurement records with the same outcome. This is just like the fully
projective Zeno limit where the system remains perfectly pure in one
of the possible projections, but the detections remain randomly
distributed in time.

One might then be concerned that purity is preserved since we might be
averaging over many trajectories within this Zeno subspace. We have
neglected the small terms $\hat{\rho}_{m,n}$ ($m,n \ne 0$) which are
$\mathcal{O}(K^2/\lambda^4)$ and thus they are not correctly accounted
for by the approximation. This means that we have an
$\mathcal{O}(K^2/\lambda^4)$ error in our density matrix and the
purity given by
\begin{equation}
  \mathrm{Tr}(\hat{\rho}^2) = \mathrm{Tr}(\hat{\rho}^2_{00} + \sum_{m \ne
    0} \hat{\rho}_{0m}\hat{\rho}_{m0}) + \mathrm{Tr}(\sum_{m,n\ne0}
  \hat{\rho}_{mn} \hat{\rho}_{nm})
\end{equation}
where the second term contains the terms not accounted for by our
approximation and thus introduces an $\mathcal{O}(K^4/\lambda^8)$
error. Therefore, this discrepancy is negligible in our
approximation. The pure state predicted by $\H_\mathrm{eff}$ is only
an approximation, albeit a good one, and the real state will be mixed
to a small extent. Whilst perfect purity within the Zeno subspace
$\hat{\rho}_{00}$ is expected due to the measurement's strong
decoupling effect, the nearly perfect purity when transitions outside
the Zeno subspace are included is a non-trivial result. Similarly, in
Raman transitions the population of the neglected excited state is
also non-zero, but negligible. Furthermore, this equation does not
actually require the adiabatic elimination used in
Eq. \eqref{eq:intermediate} (note that we only used it to convince
ourselves that the coherences are small) and such situations may be
considered provided all approximations remain valid. In a similar way
the limit of linear optics is derived from the physics of a two-level
nonlinear medium, when the population of the upper state is neglected
and the adiabatic elimination of coherences is not required.

\section{Non-Hermitian Dynamics in Ultracold Gases}

\subsection{Theoretical Model}

\begin{figure}[hbtp!]
	\includegraphics[width=\linewidth]{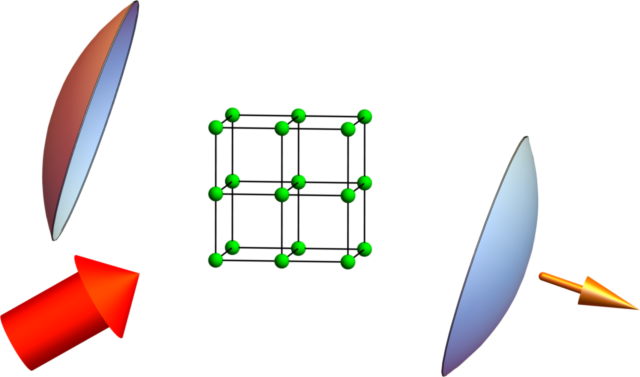}
	\caption{(Color online) Atoms in an optical lattice are probed
          by a coherent light beam, and the light scattered at a
          particular angle is enhanced and collected by a leaky
          cavity. The photons escaping the cavity are detected,
          perturbing the atomic evolution via measurement
          backaction.\label{fig:setup}}
\end{figure}

We now focus on ultracold bosons in a lattice inside a cavity which
selects and enhances light scattered at a particular angle
\cite{Bux2013, Kessler2014, Landig2015}, as shown in
Fig. \ref{fig:setup}. One of its key advantages is the flexibility of
engineering the measurement $\c$ and the possibility of coupling to
both density and inter-site interference \cite{Elliott2015,
  Kozlowski2015, Mazzucchi2015, Caballero2015, Caballero2015a}. The
global nature of such measurement is also in contrast to spontaneous
emission \cite{Pichler2010, Sarkar2014}, local \cite{Syassen2008,
  Kepesidis2012, Vidanovic2014, Bernier2014, Daley2014} and
fixed-range addressing \cite{Ates2012, Everest2014} which are
typically considered in dissipative systems. Furthermore, it provides
the opportunity to extend quantum measurement and quantum Zeno
dynamics beyond single atoms into strongly correlated many-body
systems. However, our result can be applied to a variety of other
setups such as trapped ions \cite{Lee2014a, Lee2014b}, Rydberg-dressed
Bose-Einstein condensates \cite{Otterbach2014}, circuit QED
\cite{Bretheau2015}, or other systems under continuous observation.

The isolated system is described by the Bose-Hubbard Model (BHM),
\begin{equation}
\hat{H}_0 = -J \sum_{\langle i, j \rangle} \bd_i b_j + \frac{U}{2}
\sum_i \n_i (\n_i - 1),
\end{equation}
where $b_i$ ($\bd_i$) are the bosonic annihilation (creation)
operators at site $i$, $\n_i$ is the atom number operator at site $i$,
$U$ is the on-site interaction, and $J$ is the tunnelling
coefficient. The optical Hamiltonian is adiabatically eliminated and
if the cavity field backaction can be neglected (cavity detuning must
be small compared to its decay rate) then its only effect on the
system is via measurement backaction \cite{Mekhov2012}. The quantum
jump operator is given by $\c = \sqrt{2 \kappa} \a$, where $\kappa$ is
the cavity relaxation rate, $\a = C \hat{D}$ is the annihilation
operator of a photon in the cavity mode, $C$ is the coefficient of
Rayleigh scattering into the cavity, $\hat{D} = \sum_i A_i \n_i$, with
the coefficients given by
$A_i = u^*_\mathrm{out}({\bf r}_i) u_\mathrm{in}({\bf r}_i)$, and
$u_\mathrm{in,out}({\bf r}_i)$ are the mode functions of the incoming
and scattered light \cite{Mekhov2012}. Additionally, we have the
necessary condition to be in the quantum Zeno regime
$\gamma/ J \gg 1$, where $\gamma = \kappa |C|^2$.

We will now consider the simplest case of global multi-site
measurement of the form $\hat{D} = \hat{N}_K = \sum_i^K \n_i$, where
the sum is over $K$ illuminated sites. Physically, this can be
realised by collecting the light scattered into a diffraction maximum
\cite{Mekhov2007, Mekhov2007PRA}. The effective Hamiltonian becomes
\begin{equation}
\label{eq:nHH2}
\hat{H}_\mathrm{eff} = \hat{H}_0 - i \gamma \left(  \delta \hat{N}_K \right)^2,
\end{equation}
where $ \delta \hat{N}_K = \hat{N}_K - N^0_K$ and $N^0_K$ is a
subspace eigenvalue. It is now obvious that continuous measurement
squeezes the fluctuations in the measured quantity, as expected, and
that the only competing process is the system's own dynamics.

In this case, if we adiabatically eliminate the density matrix
cross-terms and substitute Eq. \eqref{eq:intermediate} into
Eq. \eqref{eq:master} for this system we obtain an effective
Hamiltonian within the Zeno subspace defined by $N_K$
\begin{equation}
  \H_\varphi = P_\varphi \left[ \H_0 - i \frac{J^2}{\gamma} \sum_{\substack{\langle i \in \varphi, j \in \varphi^\prime \rangle \\ \langle k \in \varphi^\prime, l \in \varphi \rangle}} b^\dagger_i b_j b^\dagger_k b_l \right] P_\varphi,
\end{equation}
where $\varphi = \{N_K\}$ denotes the set of states with $N_K$ atoms
in the illuminated area, $\varphi^\prime = \{N_K \pm 1\}$ denotes the
set of intermediate states and $P_\varphi$ is the projector onto
$\varphi$. We focus on the case when the second term is not only
significant, but also leads to dynamics within $\varphi$ that are not
allowed by conventional quantum Zeno dynamics accounted for by the
first term. The second term represents second-order transitions via
other subspaces which act as intermediate states much like virtual
states in optical Raman transitions. This is in contrast to the
conventional understanding of the Zeno dynamics for infinitely
frequent projective measurements (corresponding to
$\gamma \rightarrow \infty$) where such processes are forbidden
\cite{Facchi2008}. Thus, it is the weak quantum measurement that
effectively couples the states.

\subsection{Small System Example}

To get clear physical insight, we initially consider three atoms in
three sites and choose our measurement operator such that
$\hat{D} = \n_2$, i.e.~only the middle site is subject to measurement,
and the Zeno subspace defined by $n_2 = 1$. Such an illumination
pattern can be achieved with global addressing by crossing two beams
and placing the nodes at the odd sites and the antinodes at even
sites. This means that $P_\varphi \H_0 P_\varphi = 0$. However, the
first and third sites are connected via the second term. Diagonalising
the Hamiltonian reveals that out of its ten eigenvalues all but three
have a significant negative imaginary component of the order $\gamma$
which means that the corresponding eigenstates decay on a time scale
of a single quantum jump and thus quickly become negligible. The three
remaining eigenvectors are dominated by the linear superpositions of
the three Fock states $|2,1,0 \rangle$, $|1, 1, 1 \rangle$, and
$|0,1,2 \rangle$. Whilst it is not surprising that these components
are the only ones that remain as they are the only ones that actually
lie in the Zeno subspace $n_2 = 1$, it is impossible to solve the full
dynamics by just considering these Fock states alone as they are not
coupled to each other in $\hat{H}_0$. The components lying outside of
the Zeno subspace have to be included to allow intermediate steps to
occur via states that do not belong in this subspace, much like
virtual states in optical Raman transitions.

\begin{figure}[hbtp!]
	\includegraphics[width=\linewidth]{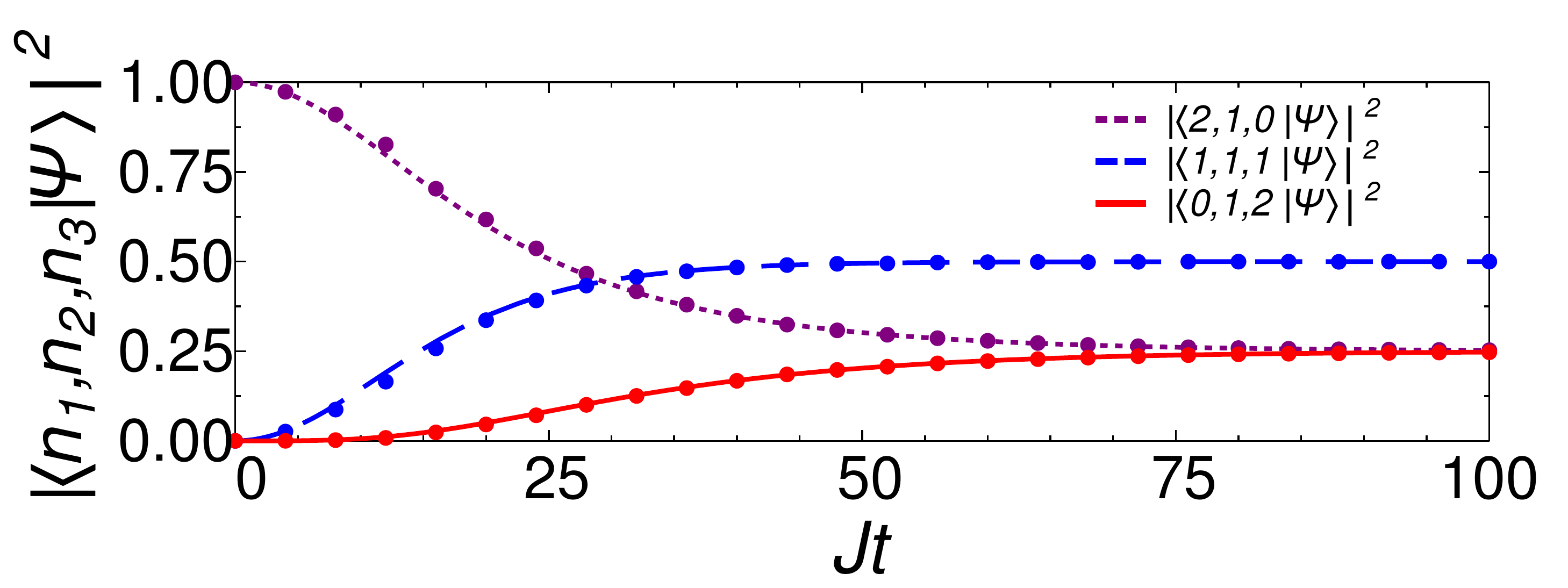}
	\caption{(Color online). Populations of the Fock states in the
          Zeno subspace for $\gamma/J = 100$ and initial state
          $| 2,1,0 \rangle$. It is clear that quantum Zeno dynamics
          occurs via Raman-like processes even though none of these
          states are connected in $\hat{H}_0$. The dynamics occurs via
          virtual intermediate states outside the Zeno subspace. The
          system also tends to a steady state which minimises
          tunnelling effectively suppressing fluctuations. The lines
          are solutions to the non-Hermitian Hamiltonian, and the dots
          are points from a stochastic trajectory
          calculation.\label{fig:comp}}
\end{figure}

An approximate solution for $U=0$ can be written for the
$\{|2,1,0 \rangle, |1,1,1 \rangle, |0,1,2 \rangle\}$ subspace by
multiplying each eigenvector with its corresponding time evolution
\begin{equation}
  | \Psi(t) \rangle \propto \left( \begin{array}{c} 
  z_1 + \sqrt{2} z_2 e^{-6 J^2 t / \gamma} + z_3 e^{-12 J^2 t / \gamma} \\
  -\sqrt{2} \left(z_1 - z_3 e^{-12 J^2 t / \gamma} \right) \\ 
  z_1 - \sqrt{2} z_2 e^{-6 J^2 t / \gamma} + z_3 e^{-12 J^2 t /
                                     \gamma} \\ 
                                   \end{array} 
                                 \right), \nonumber
\end{equation}
where $z_i$ denote the overlap between the eigenvectors and the
initial state, $z_i = \langle v_i | \Psi (0) \rangle$, with
$| v_1 \rangle = (1, -\sqrt{2}, 1)/2$,
$| v_2 \rangle = (1, 0, -1)/\sqrt{2}$, and
$| v_3 \rangle = (1, \sqrt{2}, 1)/2$. The steady state as
$t \rightarrow \infty$ is given by
$| v_1 \rangle = (1, -\sqrt{2}, 1)/2$. This solution is illustrated in
Fig. \ref{fig:comp} which clearly demonstrates dynamics beyond the
canonical understanding of quantum Zeno dynamics as tunnelling occurs
between states coupled via a different Zeno subspace.

\subsection{Steady State of non-Hermitian Dynamics}

A distinctive difference between BHM ground states and the final
steady state,
$[|2,1,0 \rangle - \sqrt{2} |1,1,1\rangle + |0,1,2\rangle]$, is that
its components are not in phase. Squeezing due to measurement
naturally competes with inter-site tunnelling which tends to spread
the atoms. However, from Eq. \eqref{eq:nHH2} we see the final state
will always be the eigenvector with the smallest fluctuations as it
will have an eigenvalue with the largest imaginary component. This
naturally corresponds to the state where tunnelling between Zeno
subspaces (here between every site) is minimised by destructive
matter-wave interference, i.e.~the tunnelling dark state defined by
$\hat{T} |\Psi \rangle = 0$, where
$\hat{T} = \sum_{\langle i, j \rangle} \bd_i b_j$. This is simply the
physical interpretation of the steady states we predicted for
\mbox{Eq. \eqref{eq:master}}. Crucially, this state can only be
reached if the dynamics aren't fully suppressed by measurement and
thus, counter-intuitively, the atomic dynamics cooperate with
measurement to suppress itself by destructive interference. Therefore,
this effect is beyond the scope of traditional quantum Zeno dynamics
and presents a new perspective on the competition between a system's
short-range dynamics and global measurement backaction.

We now consider a one-dimensional lattice with $M$ sites so we extend
the measurement to $\hat{D} = \N_\text{even}$ where every even site is
illuminated (obtained by crossing two beams such that the nodes
coincide with odd sites and antinodes with even sites
\cite{Mekhov2007PRA, Mazzucchi2015}). The wavefunction in a Zeno
subspace must be an eigenstate of $\c$ and we combine this with the
requirement for it to be in the dark state of the tunnelling operator
(eigenstate of $\H_0$ for $U = 0$) to derive the steady state. These
two conditions in momentum space are
\begin{equation}
  \hat{T} | \Psi \rangle = \sum_{\text{RBZ}} \left[ \bd_k b_k - \bd_{q} b_{q} \right] \cos(ka) |\Psi \rangle = 0, \nonumber
\end{equation}
\begin{equation}
  \Delta \N |\Psi \rangle = \sum_{\text{RBZ}} \left[ \bd_k b_{-q} + \bd_{-q} b_k \right] | \Psi \rangle= \Delta N |\Psi \rangle, \nonumber
\end{equation}
where $b_k = \frac{1}{\sqrt{M}} \sum_j e^{i k j a} b_j$,
$\Delta \hat{N} = \hat{D} - N/2$, $q = \pi/a - k$, $a$ is the lattice
spacing, $N$ the total atom number, and we perform summations over the
reduced Brillouin zone (RBZ), $-\pi/2a < k \le \pi/2a$, as the
symmetries of the system are clearer this way. Now we define
\begin{equation}
\hat{\alpha}_k^\dagger = \bd_k \bd_q - \bd_{-k} \bd_{-q},
\end{equation}
\begin{equation}
\hat{\beta}_\varphi^\dagger = \bd_{\pi/2a} + \varphi \bd_{-\pi/2a},
\end{equation}
where $\varphi = \Delta N / | \Delta N |$, which create the smallest
possible states that satisfy the two equations for $\Delta N = 0$ and
$\Delta N \ne 0$ respectively. Therefore, by noting that
$\left[ \hat{T}, \hat{\alpha}_k^\dagger \right] = \left[ \Delta \N,
  \hat{\alpha}_k^\dagger \right] = \left[ \hat{T},
  \hat{\beta}_\varphi^\dagger \right] = 0$ and
$\left[ \Delta \N, \hat{\beta}_\varphi^\dagger \right] = \varphi
\hat{\beta}_\varphi^\dagger$ we can now write the equation for the
$N$-particle steady state
\begin{equation}
\label{eq:ss}
| \Psi \rangle \propto \left[ \prod_{i=1}^{(N - |\Delta N|)/2} \left( \sum_{k = 0}^{\pi/2a} \phi_{i,k} \hat{\alpha}_k^\dagger \right) \right] \left( \hat{\beta}_\varphi^\dagger \right)^{| \Delta N |} | 0 \rangle, \nonumber
\end{equation}
where $\phi_{i,k}$ are coefficients that depend on the trajectory
taken to reach this state and $|0 \rangle$ is the vacuum state defined
by $b_k |0 \rangle = 0$. Since this a dark state (an eigenstate of
$\H_0$) of the atomic dynamics, this state will remain stationary even
with measurement switched-off. Interestingly, this state is very
different from the ground states of the BHM, it is even orthogonal to
the superfluid state, and thus it cannot be obtained by cooling or
projecting from an initial ground state. The combination of tunnelling
with measurement is necessary.

\begin{figure}[hbtp!]
	\includegraphics[width=\linewidth]{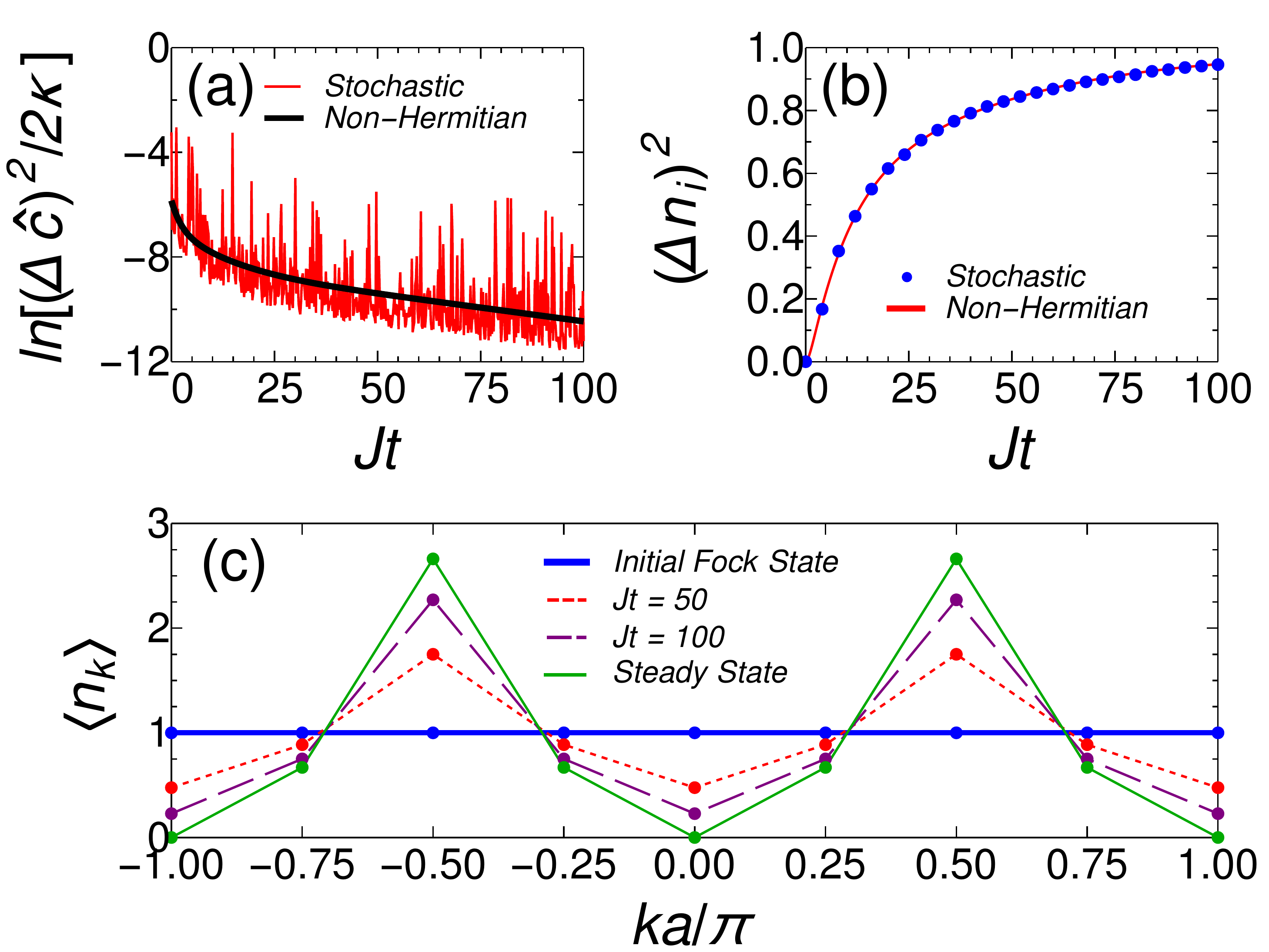}
	\caption{(Color online). A trajectory simulation for eight
          atoms in eight sites, initially in
          $|1,1,1,1,1,1,1,1 \rangle$, with periodic boundary
          conditions and $\gamma/J = 100$. (a), The fluctuations in
          $\c$ where the stochastic nature of the process is clearly
          visible on a single trajectory level. However, the general
          trend is captured by the non-Hermitian Hamiltonian. (b), The
          local density variance. Whilst the fluctuations in the
          global measurement operator decrease, the fluctuations in
          local density increase due to tunnelling via states outside
          the Zeno subspace. (c), The momentum distribution. The
          initial Fock state has a flat distribution which with time
          approaches the steady state distribution of two identical
          and symmetric distributions centred at $k = \pi/2a$ and
          $k = -\pi/2a$.\label{fig:steady}}
\end{figure}

In order to prepare the steady state one has to run the experiment and
wait until the photocount rate remains constant for a sufficiently
long time. Such a trajectory is illustrated in Fig. \ref{fig:steady}
and compared to a deterministic trajectory calculated using the
non-Hermitian Hamiltonian. It is easy to see from
Fig. \ref{fig:steady}(a) how the stochastic fluctuations around the
mean value of the observable have no effect on the general behaviour
of the system in the strong measurement regime. By discarding these
fluctuations we no longer describe a pure state, but we showed how
this only leads to a negligible error. Fig. \ref{fig:steady}(b) shows
the local density variance in the lattice. Not only does it grow
showing evidence of tunnelling between illuminated and non-illuminated
sites, but it grows to significant values. This is in contrast to
conventional quantum Zeno dynamics where no tunnelling would be
allowed at all. Finally, Fig. \ref{fig:steady}(c) shows the momentum
distribution of the trajectory. We can clearly see that it deviates
significantly from the initial flat distribution of the Fock
state. Furthermore, the steady state does not have any atoms in the
$k=0$ state and thus is orthogonal to the superfluid state as
discussed.

To obtain a state with a specific value of $\Delta N$ postselection
may be necessary, but otherwise it is not needed.  The process can be
optimised by feedback control since the state is monitored at all
times \cite{Ivanov2014}. Furthermore, the form of the measurement
operator is very flexible and it can easily be engineered by the
geometry of the optical setup \cite{Elliott2015, Mazzucchi2015} which
can be used to design a state with desired properties.

\section{Summary}

In summary, we have presented a new perspective on quantum Zeno
dynamics when the measurement isn't fully projective. By using the
fact that the system is strongly confined to a specific measurement
eigenspace we have derived an effective non-Hermitian Hamiltonian. In
contrast to previous works, it is independent of the underlying system
and there is no need to postselect for a particular exotic trajectory
\cite{Lee2014a, Lee2014b}. Using the BHM as an example we have shown
that whilst the system remains in its Zeno subspace, it will exhibit
Raman-like transitions within this subspace which would be forbidden in
the canonical fully projective limit. Finally, we have shown that the
system will always tend towards the eigenstate of the Hamiltonian with
the best squeezing in the measured quantity and the atomic dynamics,
which normally tend to spread the distribution, cooperates with
measurement to produce a state in which tunnelling is suppressed by
destructive matter-wave interference. A dark state of the tunnelling
operator will have zero fluctuations and we provided an expression for
the steady state which is significantly different from the ground
state of the Hamiltonian. This is in contrast to previous works on
dissipative state preparation where the steady state had to be a dark
state of the measurement operator instead \cite{Diehl2008}.

\begin{acknowledgments} 
The authors are grateful to EPSRC (DTA and EP/I004394/1). 
\end{acknowledgments}

\appendix
\section{Determining the Zeno subspace}
\label{appendix}

Following the main text we now consider how to estimate the Zeno
subspace. Since the density matrix cross-terms are small we know
\emph{a priori} that the individual wavefunctions comprising the
density matrix mixture will not be coherent superpositions of
different Zeno subspaces. Therefore, each individual experiment will
at any time be predominantly in a single Zeno subspace with small
cross-terms and negligible occupations in the other subspaces. With no
measurement record our density matrix would be a mixture of all these
possibilities. However, we can try and determine the Zeno subspace
around which the state evolves in a single experiment from the number
of detections, $m$, in time $t$.

The detection distribution on time-scales shorter than dissipation (so
we can approximate as if we were in a fully Zeno regime) can be
obtained by integrating over the detection times \cite{Mekhov2009PRA}
to get
\begin{equation}
  P(m,t) = \sum_n \frac{[|c_n|^2 t]^m} {m!} e^{-|c_n|^2 t} \mathrm{Tr} (\rho_{nn}).
\end{equation}
For a state that is predominantly in one Zeno subspace, the
distribution will be approximately Poissonian (up to
$\mathcal{O}(K^2 / \lambda^4$), the population of the other
subspaces). Therefore, in a single experiment we will measure
$m = |c_0|^2t \pm \sqrt{|c_0|^2t}$ detections (note, we have assumed
$|c_0|^2 t$ is large enough to approximate the distribution as
normal. This is not necessary, we simply use it here to not have to
worry about the asymmetry in the deviation around the mean value). The
uncertainty does not come from the fact that $\lambda$ is not
infinite. The jumps are random events with a Poisson
distribution. Therefore, even in the full projective limit we will not
observe the same detection trajectory in each experiment even though
the system evolves in exactly the same way and remains in a perfectly
pure state.

If the basis of $\c$ is continuous (e.g. free particle position or
momentum) then the deviation around the mean will be our upper bound
on the deviation of the system from a pure state evolving around a
single Zeno subspace. However, continuous systems are beyond the scope
of this work and we will confine ourselves to discrete systems. Though
it is important to remember that continuous systems can be treated
this way, but the error estimate (and thus the mixedness of the state)
will be different.

For a discrete system it is easier to exclude all possibilities except
for one. The error in our estimate of $|c_0|^2$ in a single experiment
decreases as $1/\sqrt{t}$ and thus it can take a long time to
confidently determine $|c_0|^2$ to a sufficient precision this
way. However, since we know that it can only take one of the possible
values from the set $\{|c_n|^2 \}$ it is much easier to exclude all
the other values.

In an experiment we can use Bayes' theorem to infer the state of our
system as follows
\begin{equation}
	p(c_n = c_0 | m) = \frac{ p(m | c_n = c_0) p(c_n = c_0) }{ p(m) },
\end{equation}
where $p(x)$ denotes the probability of the discrete event $x$ and
$p(x|y)$ the conditional probability of $x$ given $y$. We know that
$p(m | c_n = c_0)$ is simply given by a Poisson distribution with mean
$|c_0|^2 t$. $p(m)$ is just a normalising factor and $p(c_n = c_0)$ is
our \emph{a priori} knowledge of the state. Therefore, one can get the
probability of being in the right Zeno subspace from
\begin{align}
  p(c_n & = c_0 | m) = \frac{ p_0(c_n = c_0) \frac{ \left( |c_0|^2 t \right)^{2m} } {m!} e^{-|c_0|^2 t}} {\sum_n p_0(c_n) \frac{ \left( |c_n|^2 t \right)^{2m} } {m!} e^{-|c_n|^2 t}} \nonumber \\
	& = p_0(c_n = c_0) \left[ \sum_n p_0(c_n) \left( \frac{ |c_n|^2 } { |c_0|^2 } \right)^{2m} e^{\left( |c_0|^2 - |c_n|^2 \right) t} \right]^{-1},
\end{align}
where $p_0$ denotes probabilities at $t = 0$. In a real experiment one
could prepare the initial state to be close to the Zeno subspace of
interest and thus it would be easier to deduce the state. Furthermore,
in the middle of an experiment if we have already established the Zeno
subspace this will be reflected in these \emph{a priori} probabilities
again making it easier to infer the correct subspace. However, we will
consider the worst case scenario which might be useful if we don't
know the initial state or if the Zeno subspace changes during the
experiment, a uniform $p_0(c_n)$.

This probability is a rather complicated function as $m$ is a
stochastic quantity that also increases with $t$. We want it to be as
close to 1 as possible. In order to devise an appropriate condition
for this we note that in the first line all terms in the denominator
are Poisson distributions of $m$. Therefore, if the mean values
$|c_n|^2 t$ are sufficiently spaced out, only one of the terms in the
sum will be significant for a given $m$ and if this happens to be the
one that corresponds to $c_0$ we get a probability close to
unity. Therefore, we set the condition such that it is highly unlikely
that our measured $m$ could be produced by two different distributions
\begin{align}
  \sqrt{|c_0|^2 t} \ll ||c_0|^2 - |c_n|^2|| t, \forall n \ne 0 \\
  \sqrt{|c_n|^2 t} \ll ||c_0|^2 - |c_n|^2|| t, \forall n \ne 0
\end{align}
The LHS is the standard deviation of $m$ if the system was in subspace
$P_0$ or $P_n$. The RHS is the difference in the mean detections
between the subspace $n$ and the one we are interested in. The
condition becomes more strict if the subspaces become less
distinguishable as it becomes harder to confidently determine the
correct state. Once again, using $\c = \lambda \hat{o}$ where
$\hat{o} \sim \mathcal{O}(1)$ we get
\begin{equation}
  t \gg \frac{1}{\lambda^2} \frac{|o_{0,n}|^2} {(|o_0|^2 - |o_n|^2|)^2}.
\end{equation}
Since detections happen on average at an average rate of order
$\lambda^2$ we only need to wait for a few detections to satisfy this
condition. Therefore, we see that even in the worst case scenario of
complete ignorance of the state of the system we can very easily
determine the correct subspace. Once it is established for the first
time, the \emph{a priori} information can be updated and it will
become even easier to monitor the system.

However, it is important to note that physically once the quantum
jumps deviate too much from the mean value the system is more likely
to change the Zeno subspace (due to measurement backaction) and the
detection rate will visibly change. Therefore, if we observe a
consistent detection rate it is extremely unlikely that it can be
produced by two different Zeno subspaces so in fact it is even easier
to determine the correct state, but the above estimate serves as a
good lower bound on the necessary detection time.

\bibliographystyle{apsrev4-1}
\bibliography{manuscript}

\begin{thebibliography}{88}%
\makeatletter
\providecommand \@ifxundefined [1]{%
 \@ifx{#1\undefined}
}%
\providecommand \@ifnum [1]{%
 \ifnum #1\expandafter \@firstoftwo
 \else \expandafter \@secondoftwo
 \fi
}%
\providecommand \@ifx [1]{%
 \ifx #1\expandafter \@firstoftwo
 \else \expandafter \@secondoftwo
 \fi
}%
\providecommand \natexlab [1]{#1}%
\providecommand \enquote  [1]{``#1''}%
\providecommand \bibnamefont  [1]{#1}%
\providecommand \bibfnamefont [1]{#1}%
\providecommand \citenamefont [1]{#1}%
\providecommand \href@noop [0]{\@secondoftwo}%
\providecommand \href [0]{\begingroup \@sanitize@url \@href}%
\providecommand \@href[1]{\@@startlink{#1}\@@href}%
\providecommand \@@href[1]{\endgroup#1\@@endlink}%
\providecommand \@sanitize@url [0]{\catcode `\\12\catcode `\$12\catcode
  `\&12\catcode `\#12\catcode `\^12\catcode `\_12\catcode `\%12\relax}%
\providecommand \@@startlink[1]{}%
\providecommand \@@endlink[0]{}%
\providecommand \url  [0]{\begingroup\@sanitize@url \@url }%
\providecommand \@url [1]{\endgroup\@href {#1}{\urlprefix }}%
\providecommand \urlprefix  [0]{URL }%
\providecommand \Eprint [0]{\href }%
\providecommand \doibase [0]{http://dx.doi.org/}%
\providecommand \selectlanguage [0]{\@gobble}%
\providecommand \bibinfo  [0]{\@secondoftwo}%
\providecommand \bibfield  [0]{\@secondoftwo}%
\providecommand \translation [1]{[#1]}%
\providecommand \BibitemOpen [0]{}%
\providecommand \bibitemStop [0]{}%
\providecommand \bibitemNoStop [0]{.\EOS\space}%
\providecommand \EOS [0]{\spacefactor3000\relax}%
\providecommand \BibitemShut  [1]{\csname bibitem#1\endcsname}%
\let\auto@bib@innerbib\@empty
\bibitem [{\citenamefont {Misra}\ and\ \citenamefont
  {Sudarshan}(1977)}]{Zeno1977}%
  \BibitemOpen
  \bibfield  {author} {\bibinfo {author} {\bibfnamefont {B.}~\bibnamefont
  {Misra}}\ and\ \bibinfo {author} {\bibfnamefont {E.~C.~G.}\ \bibnamefont
  {Sudarshan}},\ }\href@noop {} {\bibfield  {journal} {\bibinfo  {journal} {J.
  Math. Phys.}\ }\textbf {\bibinfo {volume} {18}},\ \bibinfo {pages} {756}
  (\bibinfo {year} {1977})}\BibitemShut {NoStop}%
\bibitem [{\citenamefont {Facchi}\ and\ \citenamefont
  {Pascazio}(2008)}]{Facchi2008}%
  \BibitemOpen
  \bibfield  {author} {\bibinfo {author} {\bibfnamefont {P.}~\bibnamefont
  {Facchi}}\ and\ \bibinfo {author} {\bibfnamefont {S.}~\bibnamefont
  {Pascazio}},\ }\href@noop {} {\bibfield  {journal} {\bibinfo  {journal} {J.
  Phys. A: Math. Theor.}\ }\textbf {\bibinfo {volume} {41}},\ \bibinfo {pages}
  {493001} (\bibinfo {year} {2008})}\BibitemShut {NoStop}%
\bibitem [{\citenamefont {Itano}\ \emph {et~al.}(1990)\citenamefont {Itano},
  \citenamefont {Heinzen}, \citenamefont {Bollinger},\ and\ \citenamefont
  {Wineland}}]{Itano1990}%
  \BibitemOpen
  \bibfield  {author} {\bibinfo {author} {\bibfnamefont {W.~M.}\ \bibnamefont
  {Itano}}, \bibinfo {author} {\bibfnamefont {D.~J.}\ \bibnamefont {Heinzen}},
  \bibinfo {author} {\bibfnamefont {J.~J.}\ \bibnamefont {Bollinger}}, \ and\
  \bibinfo {author} {\bibfnamefont {D.~J.}\ \bibnamefont {Wineland}},\
  }\href@noop {} {\bibfield  {journal} {\bibinfo  {journal} {Phys. Rev. A}\
  }\textbf {\bibinfo {volume} {41}},\ \bibinfo {pages} {2295} (\bibinfo {year}
  {1990})}\BibitemShut {NoStop}%
\bibitem [{\citenamefont {Nagels}\ \emph {et~al.}(1997)\citenamefont {Nagels},
  \citenamefont {Hermans},\ and\ \citenamefont {Chapovsky}}]{Nagels1997}%
  \BibitemOpen
  \bibfield  {author} {\bibinfo {author} {\bibfnamefont {B.}~\bibnamefont
  {Nagels}}, \bibinfo {author} {\bibfnamefont {L.~J.~F.}\ \bibnamefont
  {Hermans}}, \ and\ \bibinfo {author} {\bibfnamefont {P.~L.}\ \bibnamefont
  {Chapovsky}},\ }\href@noop {} {\bibfield  {journal} {\bibinfo  {journal}
  {Phys. Rev. Lett.}\ }\textbf {\bibinfo {volume} {79}},\ \bibinfo {pages}
  {3097} (\bibinfo {year} {1997})}\BibitemShut {NoStop}%
\bibitem [{\citenamefont {Kwiat}\ \emph {et~al.}(1999)\citenamefont {Kwiat},
  \citenamefont {White}, \citenamefont {Mitchell}, \citenamefont {Nairz},
  \citenamefont {Weihs}, \citenamefont {Weinfurter},\ and\ \citenamefont
  {Zeilinger}}]{Kwiat1999}%
  \BibitemOpen
  \bibfield  {author} {\bibinfo {author} {\bibfnamefont {P.~G.}\ \bibnamefont
  {Kwiat}}, \bibinfo {author} {\bibfnamefont {A.~G.}\ \bibnamefont {White}},
  \bibinfo {author} {\bibfnamefont {J.~R.}\ \bibnamefont {Mitchell}}, \bibinfo
  {author} {\bibfnamefont {O.}~\bibnamefont {Nairz}}, \bibinfo {author}
  {\bibfnamefont {G.}~\bibnamefont {Weihs}}, \bibinfo {author} {\bibfnamefont
  {H.}~\bibnamefont {Weinfurter}}, \ and\ \bibinfo {author} {\bibfnamefont
  {A.}~\bibnamefont {Zeilinger}},\ }\href@noop {} {\bibfield  {journal}
  {\bibinfo  {journal} {Phys. Rev. Lett.}\ }\textbf {\bibinfo {volume} {83}},\
  \bibinfo {pages} {4725} (\bibinfo {year} {1999})}\BibitemShut {NoStop}%
\bibitem [{\citenamefont {Balzer}\ \emph {et~al.}(2000)\citenamefont {Balzer},
  \citenamefont {Huesmann}, \citenamefont {Neuhauser},\ and\ \citenamefont
  {Toschek}}]{Balzer2000}%
  \BibitemOpen
  \bibfield  {author} {\bibinfo {author} {\bibfnamefont {C.}~\bibnamefont
  {Balzer}}, \bibinfo {author} {\bibfnamefont {R.}~\bibnamefont {Huesmann}},
  \bibinfo {author} {\bibfnamefont {W.}~\bibnamefont {Neuhauser}}, \ and\
  \bibinfo {author} {\bibfnamefont {P.}~\bibnamefont {Toschek}},\ }\href@noop
  {} {\bibfield  {journal} {\bibinfo  {journal} {Optics Communications}\
  }\textbf {\bibinfo {volume} {180}},\ \bibinfo {pages} {115} (\bibinfo {year}
  {2000})}\BibitemShut {NoStop}%
\bibitem [{\citenamefont {Streed}\ \emph {et~al.}(2006)\citenamefont {Streed},
  \citenamefont {Mun}, \citenamefont {Boyd}, \citenamefont {Campbell},
  \citenamefont {Medley}, \citenamefont {Ketterle},\ and\ \citenamefont
  {Pritchard}}]{Streed2006}%
  \BibitemOpen
  \bibfield  {author} {\bibinfo {author} {\bibfnamefont {E.~W.}\ \bibnamefont
  {Streed}}, \bibinfo {author} {\bibfnamefont {J.}~\bibnamefont {Mun}},
  \bibinfo {author} {\bibfnamefont {M.}~\bibnamefont {Boyd}}, \bibinfo {author}
  {\bibfnamefont {G.~K.}\ \bibnamefont {Campbell}}, \bibinfo {author}
  {\bibfnamefont {P.}~\bibnamefont {Medley}}, \bibinfo {author} {\bibfnamefont
  {W.}~\bibnamefont {Ketterle}}, \ and\ \bibinfo {author} {\bibfnamefont
  {D.~E.}\ \bibnamefont {Pritchard}},\ }\href@noop {} {\bibfield  {journal}
  {\bibinfo  {journal} {Phys. Rev. Lett.}\ }\textbf {\bibinfo {volume} {97}},\
  \bibinfo {pages} {260402} (\bibinfo {year} {2006})}\BibitemShut {NoStop}%
\bibitem [{\citenamefont {Hosten}\ \emph {et~al.}(2006)\citenamefont {Hosten},
  \citenamefont {Rakher}, \citenamefont {Barreiro}, \citenamefont {Peters},\
  and\ \citenamefont {Kwiat}}]{Hosten2006}%
  \BibitemOpen
  \bibfield  {author} {\bibinfo {author} {\bibfnamefont {O.}~\bibnamefont
  {Hosten}}, \bibinfo {author} {\bibfnamefont {M.~T.}\ \bibnamefont {Rakher}},
  \bibinfo {author} {\bibfnamefont {J.~T.}\ \bibnamefont {Barreiro}}, \bibinfo
  {author} {\bibfnamefont {N.~A.}\ \bibnamefont {Peters}}, \ and\ \bibinfo
  {author} {\bibfnamefont {P.~G.}\ \bibnamefont {Kwiat}},\ }\href@noop {}
  {\bibfield  {journal} {\bibinfo  {journal} {Nature}\ }\textbf {\bibinfo
  {volume} {439}},\ \bibinfo {pages} {949} (\bibinfo {year}
  {2006})}\BibitemShut {NoStop}%
\bibitem [{\citenamefont {Bernu}\ \emph {et~al.}(2008)\citenamefont {Bernu},
  \citenamefont {Del\'eglise}, \citenamefont {Sayrin}, \citenamefont {Kuhr},
  \citenamefont {Dotsenko}, \citenamefont {Brune}, \citenamefont {Raimond},\
  and\ \citenamefont {Haroche}}]{Bernu2008}%
  \BibitemOpen
  \bibfield  {author} {\bibinfo {author} {\bibfnamefont {J.}~\bibnamefont
  {Bernu}}, \bibinfo {author} {\bibfnamefont {S.}~\bibnamefont {Del\'eglise}},
  \bibinfo {author} {\bibfnamefont {C.}~\bibnamefont {Sayrin}}, \bibinfo
  {author} {\bibfnamefont {S.}~\bibnamefont {Kuhr}}, \bibinfo {author}
  {\bibfnamefont {I.}~\bibnamefont {Dotsenko}}, \bibinfo {author}
  {\bibfnamefont {M.}~\bibnamefont {Brune}}, \bibinfo {author} {\bibfnamefont
  {J.~M.}\ \bibnamefont {Raimond}}, \ and\ \bibinfo {author} {\bibfnamefont
  {S.}~\bibnamefont {Haroche}},\ }\href@noop {} {\bibfield  {journal} {\bibinfo
   {journal} {Phys. Rev. Lett.}\ }\textbf {\bibinfo {volume} {101}},\ \bibinfo
  {pages} {180402} (\bibinfo {year} {2008})}\BibitemShut {NoStop}%
\bibitem [{\citenamefont {Raimond}\ \emph {et~al.}(2010)\citenamefont
  {Raimond}, \citenamefont {Sayrin}, \citenamefont {Gleyzes}, \citenamefont
  {Dotsenko}, \citenamefont {Brune}, \citenamefont {Haroche}, \citenamefont
  {Facchi},\ and\ \citenamefont {Pascazio}}]{Raimond2010}%
  \BibitemOpen
  \bibfield  {author} {\bibinfo {author} {\bibfnamefont {J.~M.}\ \bibnamefont
  {Raimond}}, \bibinfo {author} {\bibfnamefont {C.}~\bibnamefont {Sayrin}},
  \bibinfo {author} {\bibfnamefont {S.}~\bibnamefont {Gleyzes}}, \bibinfo
  {author} {\bibfnamefont {I.}~\bibnamefont {Dotsenko}}, \bibinfo {author}
  {\bibfnamefont {M.}~\bibnamefont {Brune}}, \bibinfo {author} {\bibfnamefont
  {S.}~\bibnamefont {Haroche}}, \bibinfo {author} {\bibfnamefont
  {P.}~\bibnamefont {Facchi}}, \ and\ \bibinfo {author} {\bibfnamefont
  {S.}~\bibnamefont {Pascazio}},\ }\href@noop {} {\bibfield  {journal}
  {\bibinfo  {journal} {Phys. Rev. Lett.}\ }\textbf {\bibinfo {volume} {105}},\
  \bibinfo {pages} {213601} (\bibinfo {year} {2010})}\BibitemShut {NoStop}%
\bibitem [{\citenamefont {Raimond}\ \emph {et~al.}(2012)\citenamefont
  {Raimond}, \citenamefont {Facchi}, \citenamefont {Peaudecerf}, \citenamefont
  {Pascazio}, \citenamefont {Sayrin}, \citenamefont {Dotsenko}, \citenamefont
  {Gleyzes}, \citenamefont {Brune},\ and\ \citenamefont
  {Haroche}}]{Raimond2012}%
  \BibitemOpen
  \bibfield  {author} {\bibinfo {author} {\bibfnamefont {J.~M.}\ \bibnamefont
  {Raimond}}, \bibinfo {author} {\bibfnamefont {P.}~\bibnamefont {Facchi}},
  \bibinfo {author} {\bibfnamefont {B.}~\bibnamefont {Peaudecerf}}, \bibinfo
  {author} {\bibfnamefont {S.}~\bibnamefont {Pascazio}}, \bibinfo {author}
  {\bibfnamefont {C.}~\bibnamefont {Sayrin}}, \bibinfo {author} {\bibfnamefont
  {I.}~\bibnamefont {Dotsenko}}, \bibinfo {author} {\bibfnamefont
  {S.}~\bibnamefont {Gleyzes}}, \bibinfo {author} {\bibfnamefont
  {M.}~\bibnamefont {Brune}}, \ and\ \bibinfo {author} {\bibfnamefont
  {S.}~\bibnamefont {Haroche}},\ }\href@noop {} {\bibfield  {journal} {\bibinfo
   {journal} {Phys. Rev. A}\ }\textbf {\bibinfo {volume} {86}},\ \bibinfo
  {pages} {032120} (\bibinfo {year} {2012})}\BibitemShut {NoStop}%
\bibitem [{\citenamefont {Signoles}\ \emph {et~al.}(2014)\citenamefont
  {Signoles}, \citenamefont {Facon}, \citenamefont {Grosso}, \citenamefont
  {Dotsenko}, \citenamefont {Haroche}, \citenamefont {Raimond}, \citenamefont
  {Brune},\ and\ \citenamefont {Gleyzes}}]{Signoles2014}%
  \BibitemOpen
  \bibfield  {author} {\bibinfo {author} {\bibfnamefont {A.}~\bibnamefont
  {Signoles}}, \bibinfo {author} {\bibfnamefont {A.}~\bibnamefont {Facon}},
  \bibinfo {author} {\bibfnamefont {D.}~\bibnamefont {Grosso}}, \bibinfo
  {author} {\bibfnamefont {I.}~\bibnamefont {Dotsenko}}, \bibinfo {author}
  {\bibfnamefont {S.}~\bibnamefont {Haroche}}, \bibinfo {author} {\bibfnamefont
  {J.-M.}\ \bibnamefont {Raimond}}, \bibinfo {author} {\bibfnamefont
  {M.}~\bibnamefont {Brune}}, \ and\ \bibinfo {author} {\bibfnamefont
  {S.}~\bibnamefont {Gleyzes}},\ }\href@noop {} {\bibfield  {journal} {\bibinfo
   {journal} {Nat. Phys.}\ }\textbf {\bibinfo {volume} {10}},\ \bibinfo {pages}
  {715} (\bibinfo {year} {2014})}\BibitemShut {NoStop}%
\bibitem [{\citenamefont {Hatano}\ and\ \citenamefont
  {Nelson}(1996)}]{Hatano1996}%
  \BibitemOpen
  \bibfield  {author} {\bibinfo {author} {\bibfnamefont {N.}~\bibnamefont
  {Hatano}}\ and\ \bibinfo {author} {\bibfnamefont {D.~R.}\ \bibnamefont
  {Nelson}},\ }\href@noop {} {\bibfield  {journal} {\bibinfo  {journal} {Phys.
  Rev. Lett.}\ }\textbf {\bibinfo {volume} {77}},\ \bibinfo {pages} {570}
  (\bibinfo {year} {1996})}\BibitemShut {NoStop}%
\bibitem [{\citenamefont {Refael}\ \emph {et~al.}(2006)\citenamefont {Refael},
  \citenamefont {Hofstetter},\ and\ \citenamefont {Nelson}}]{Refael2006}%
  \BibitemOpen
  \bibfield  {author} {\bibinfo {author} {\bibfnamefont {G.}~\bibnamefont
  {Refael}}, \bibinfo {author} {\bibfnamefont {W.}~\bibnamefont {Hofstetter}},
  \ and\ \bibinfo {author} {\bibfnamefont {D.~R.}\ \bibnamefont {Nelson}},\
  }\href@noop {} {\bibfield  {journal} {\bibinfo  {journal} {Phys. Rev. B}\
  }\textbf {\bibinfo {volume} {74}},\ \bibinfo {pages} {174520} (\bibinfo
  {year} {2006})}\BibitemShut {NoStop}%
\bibitem [{\citenamefont {Bender}\ and\ \citenamefont
  {Boettcher}(1998)}]{Bender1998}%
  \BibitemOpen
  \bibfield  {author} {\bibinfo {author} {\bibfnamefont {C.~M.}\ \bibnamefont
  {Bender}}\ and\ \bibinfo {author} {\bibfnamefont {S.}~\bibnamefont
  {Boettcher}},\ }\href@noop {} {\bibfield  {journal} {\bibinfo  {journal}
  {Phys. Rev. Lett.}\ }\textbf {\bibinfo {volume} {80}},\ \bibinfo {pages}
  {5243} (\bibinfo {year} {1998})}\BibitemShut {NoStop}%
\bibitem [{\citenamefont {Giorgi}(2010)}]{Giorgi2010}%
  \BibitemOpen
  \bibfield  {author} {\bibinfo {author} {\bibfnamefont {G.~L.}\ \bibnamefont
  {Giorgi}},\ }\href@noop {} {\bibfield  {journal} {\bibinfo  {journal} {Phys.
  Rev. B}\ }\textbf {\bibinfo {volume} {82}},\ \bibinfo {pages} {052404}
  (\bibinfo {year} {2010})}\BibitemShut {NoStop}%
\bibitem [{\citenamefont {Zhang}\ and\ \citenamefont {Song}(2013)}]{Zhang2013}%
  \BibitemOpen
  \bibfield  {author} {\bibinfo {author} {\bibfnamefont {X.~Z.}\ \bibnamefont
  {Zhang}}\ and\ \bibinfo {author} {\bibfnamefont {Z.}~\bibnamefont {Song}},\
  }\href@noop {} {\bibfield  {journal} {\bibinfo  {journal} {Phys. Rev. A}\
  }\textbf {\bibinfo {volume} {87}},\ \bibinfo {pages} {012114} (\bibinfo
  {year} {2013})}\BibitemShut {NoStop}%
\bibitem [{\citenamefont {Otterbach}\ and\ \citenamefont
  {Lemeshko}(2014)}]{Otterbach2014}%
  \BibitemOpen
  \bibfield  {author} {\bibinfo {author} {\bibfnamefont {J.}~\bibnamefont
  {Otterbach}}\ and\ \bibinfo {author} {\bibfnamefont {M.}~\bibnamefont
  {Lemeshko}},\ }\href@noop {} {\bibfield  {journal} {\bibinfo  {journal}
  {Phys. Rev. Lett.}\ }\textbf {\bibinfo {volume} {113}},\ \bibinfo {pages}
  {070401} (\bibinfo {year} {2014})}\BibitemShut {NoStop}%
\bibitem [{\citenamefont {Lee}\ and\ \citenamefont {Chan}(2014)}]{Lee2014a}%
  \BibitemOpen
  \bibfield  {author} {\bibinfo {author} {\bibfnamefont {T.~E.}\ \bibnamefont
  {Lee}}\ and\ \bibinfo {author} {\bibfnamefont {C.-K.}\ \bibnamefont {Chan}},\
  }\href@noop {} {\bibfield  {journal} {\bibinfo  {journal} {Phys. Rev. X}\
  }\textbf {\bibinfo {volume} {4}},\ \bibinfo {pages} {041001} (\bibinfo {year}
  {2014})}\BibitemShut {NoStop}%
\bibitem [{\citenamefont {Lee}\ \emph {et~al.}(2014)\citenamefont {Lee},
  \citenamefont {Reiter},\ and\ \citenamefont {Moiseyev}}]{Lee2014b}%
  \BibitemOpen
  \bibfield  {author} {\bibinfo {author} {\bibfnamefont {T.~E.}\ \bibnamefont
  {Lee}}, \bibinfo {author} {\bibfnamefont {F.}~\bibnamefont {Reiter}}, \ and\
  \bibinfo {author} {\bibfnamefont {N.}~\bibnamefont {Moiseyev}},\ }\href@noop
  {} {\bibfield  {journal} {\bibinfo  {journal} {Phys. Rev. Lett.}\ }\textbf
  {\bibinfo {volume} {113}},\ \bibinfo {pages} {250401} (\bibinfo {year}
  {2014})}\BibitemShut {NoStop}%
\bibitem [{\citenamefont {Dembowski}\ \emph {et~al.}(2001)\citenamefont
  {Dembowski}, \citenamefont {Gr\"af}, \citenamefont {Harney}, \citenamefont
  {Heine}, \citenamefont {Heiss}, \citenamefont {Rehfeld},\ and\ \citenamefont
  {Richter}}]{Dembowski2001}%
  \BibitemOpen
  \bibfield  {author} {\bibinfo {author} {\bibfnamefont {C.}~\bibnamefont
  {Dembowski}}, \bibinfo {author} {\bibfnamefont {H.-D.}\ \bibnamefont
  {Gr\"af}}, \bibinfo {author} {\bibfnamefont {H.~L.}\ \bibnamefont {Harney}},
  \bibinfo {author} {\bibfnamefont {A.}~\bibnamefont {Heine}}, \bibinfo
  {author} {\bibfnamefont {W.~D.}\ \bibnamefont {Heiss}}, \bibinfo {author}
  {\bibfnamefont {H.}~\bibnamefont {Rehfeld}}, \ and\ \bibinfo {author}
  {\bibfnamefont {A.}~\bibnamefont {Richter}},\ }\href@noop {} {\bibfield
  {journal} {\bibinfo  {journal} {Phys. Rev. Lett.}\ }\textbf {\bibinfo
  {volume} {86}},\ \bibinfo {pages} {787} (\bibinfo {year} {2001})}\BibitemShut
  {NoStop}%
\bibitem [{\citenamefont {Choi}\ \emph {et~al.}(2010)\citenamefont {Choi},
  \citenamefont {Kang}, \citenamefont {Lim}, \citenamefont {Kim}, \citenamefont
  {Kim}, \citenamefont {Lee},\ and\ \citenamefont {An}}]{Choi2010}%
  \BibitemOpen
  \bibfield  {author} {\bibinfo {author} {\bibfnamefont {Y.}~\bibnamefont
  {Choi}}, \bibinfo {author} {\bibfnamefont {S.}~\bibnamefont {Kang}}, \bibinfo
  {author} {\bibfnamefont {S.}~\bibnamefont {Lim}}, \bibinfo {author}
  {\bibfnamefont {W.}~\bibnamefont {Kim}}, \bibinfo {author} {\bibfnamefont
  {J.-R.}\ \bibnamefont {Kim}}, \bibinfo {author} {\bibfnamefont {J.-H.}\
  \bibnamefont {Lee}}, \ and\ \bibinfo {author} {\bibfnamefont
  {K.}~\bibnamefont {An}},\ }\href@noop {} {\bibfield  {journal} {\bibinfo
  {journal} {Phys. Rev. Lett.}\ }\textbf {\bibinfo {volume} {104}},\ \bibinfo
  {pages} {153601} (\bibinfo {year} {2010})}\BibitemShut {NoStop}%
\bibitem [{\citenamefont {Ruter}\ \emph {et~al.}(2010)\citenamefont {Ruter},
  \citenamefont {Makris}, \citenamefont {El-Ganainy}, \citenamefont
  {Christodoulides}, \citenamefont {Segev},\ and\ \citenamefont
  {Kip}}]{Ruter2010}%
  \BibitemOpen
  \bibfield  {author} {\bibinfo {author} {\bibfnamefont {C.~E.}\ \bibnamefont
  {Ruter}}, \bibinfo {author} {\bibfnamefont {K.~G.}\ \bibnamefont {Makris}},
  \bibinfo {author} {\bibfnamefont {R.}~\bibnamefont {El-Ganainy}}, \bibinfo
  {author} {\bibfnamefont {D.~N.}\ \bibnamefont {Christodoulides}}, \bibinfo
  {author} {\bibfnamefont {M.}~\bibnamefont {Segev}}, \ and\ \bibinfo {author}
  {\bibfnamefont {D.}~\bibnamefont {Kip}},\ }\href@noop {} {\bibfield
  {journal} {\bibinfo  {journal} {Nat. Phys.}\ }\textbf {\bibinfo {volume}
  {6}},\ \bibinfo {pages} {192} (\bibinfo {year} {2010})}\BibitemShut {NoStop}%
\bibitem [{\citenamefont {Barontini}\ \emph {et~al.}(2013)\citenamefont
  {Barontini}, \citenamefont {Labouvie}, \citenamefont {Stubenrauch},
  \citenamefont {Vogler}, \citenamefont {Guarrera},\ and\ \citenamefont
  {Ott}}]{Barontini2013}%
  \BibitemOpen
  \bibfield  {author} {\bibinfo {author} {\bibfnamefont {G.}~\bibnamefont
  {Barontini}}, \bibinfo {author} {\bibfnamefont {R.}~\bibnamefont {Labouvie}},
  \bibinfo {author} {\bibfnamefont {F.}~\bibnamefont {Stubenrauch}}, \bibinfo
  {author} {\bibfnamefont {A.}~\bibnamefont {Vogler}}, \bibinfo {author}
  {\bibfnamefont {V.}~\bibnamefont {Guarrera}}, \ and\ \bibinfo {author}
  {\bibfnamefont {H.}~\bibnamefont {Ott}},\ }\href@noop {} {\bibfield
  {journal} {\bibinfo  {journal} {Phys. Rev. Lett.}\ }\textbf {\bibinfo
  {volume} {110}},\ \bibinfo {pages} {035302} (\bibinfo {year}
  {2013})}\BibitemShut {NoStop}%
\bibitem [{\citenamefont {Gao}\ \emph {et~al.}(2015)\citenamefont {Gao},
  \citenamefont {Estrecho}, \citenamefont {Bliokh}, \citenamefont {Liew},
  \citenamefont {Fraser}, \citenamefont {Brodbeck}, \citenamefont {Kamp},
  \citenamefont {Schneider}, \citenamefont {H{\"o}fling}, \citenamefont
  {Yamamoto} \emph {et~al.}}]{Gao2015}%
  \BibitemOpen
  \bibfield  {author} {\bibinfo {author} {\bibfnamefont {T.}~\bibnamefont
  {Gao}}, \bibinfo {author} {\bibfnamefont {E.}~\bibnamefont {Estrecho}},
  \bibinfo {author} {\bibfnamefont {K.}~\bibnamefont {Bliokh}}, \bibinfo
  {author} {\bibfnamefont {T.}~\bibnamefont {Liew}}, \bibinfo {author}
  {\bibfnamefont {M.}~\bibnamefont {Fraser}}, \bibinfo {author} {\bibfnamefont
  {S.}~\bibnamefont {Brodbeck}}, \bibinfo {author} {\bibfnamefont
  {M.}~\bibnamefont {Kamp}}, \bibinfo {author} {\bibfnamefont {C.}~\bibnamefont
  {Schneider}}, \bibinfo {author} {\bibfnamefont {S.}~\bibnamefont
  {H{\"o}fling}}, \bibinfo {author} {\bibfnamefont {Y.}~\bibnamefont
  {Yamamoto}},  \emph {et~al.},\ }\href@noop {} {\bibfield  {journal} {\bibinfo
   {journal} {Nature}\ }\textbf {\bibinfo {volume} {advance online
  publication}} (\bibinfo {year} {2015})}\BibitemShut {NoStop}%
\bibitem [{\citenamefont {Dhar}\ \emph {et~al.}(2015)\citenamefont {Dhar},
  \citenamefont {Dasgupta}, \citenamefont {Dhar},\ and\ \citenamefont
  {Sen}}]{Dhar2015}%
  \BibitemOpen
  \bibfield  {author} {\bibinfo {author} {\bibfnamefont {S.}~\bibnamefont
  {Dhar}}, \bibinfo {author} {\bibfnamefont {S.}~\bibnamefont {Dasgupta}},
  \bibinfo {author} {\bibfnamefont {A.}~\bibnamefont {Dhar}}, \ and\ \bibinfo
  {author} {\bibfnamefont {D.}~\bibnamefont {Sen}},\ }\href@noop {} {\bibfield
  {journal} {\bibinfo  {journal} {Phys. Rev. A}\ }\textbf {\bibinfo {volume}
  {91}},\ \bibinfo {pages} {062115} (\bibinfo {year} {2015})}\BibitemShut
  {NoStop}%
\bibitem [{\citenamefont {Stannigel}\ \emph {et~al.}(2014)\citenamefont
  {Stannigel}, \citenamefont {Hauke}, \citenamefont {Marcos}, \citenamefont
  {Hafezi}, \citenamefont {Diehl}, \citenamefont {Dalmonte},\ and\
  \citenamefont {Zoller}}]{Stannigel2014}%
  \BibitemOpen
  \bibfield  {author} {\bibinfo {author} {\bibfnamefont {K.}~\bibnamefont
  {Stannigel}}, \bibinfo {author} {\bibfnamefont {P.}~\bibnamefont {Hauke}},
  \bibinfo {author} {\bibfnamefont {D.}~\bibnamefont {Marcos}}, \bibinfo
  {author} {\bibfnamefont {M.}~\bibnamefont {Hafezi}}, \bibinfo {author}
  {\bibfnamefont {S.}~\bibnamefont {Diehl}}, \bibinfo {author} {\bibfnamefont
  {M.}~\bibnamefont {Dalmonte}}, \ and\ \bibinfo {author} {\bibfnamefont
  {P.}~\bibnamefont {Zoller}},\ }\href@noop {} {\bibfield  {journal} {\bibinfo
  {journal} {Physical review letters}\ }\textbf {\bibinfo {volume} {112}},\
  \bibinfo {pages} {120406} (\bibinfo {year} {2014})}\BibitemShut {NoStop}%
\bibitem [{\citenamefont {Bloch}\ \emph {et~al.}(2007)\citenamefont {Bloch},
  \citenamefont {Dalibard},\ and\ \citenamefont {Zwerger}}]{Bloch2007}%
  \BibitemOpen
  \bibfield  {author} {\bibinfo {author} {\bibfnamefont {I.}~\bibnamefont
  {Bloch}}, \bibinfo {author} {\bibfnamefont {J.}~\bibnamefont {Dalibard}}, \
  and\ \bibinfo {author} {\bibfnamefont {W.}~\bibnamefont {Zwerger}},\
  }\href@noop {} {\bibfield  {journal} {\bibinfo  {journal} {Rev. Mod. Phys.}\
  }\textbf {\bibinfo {volume} {80}},\ \bibinfo {pages} {885} (\bibinfo {year}
  {2007})}\BibitemShut {NoStop}%
\bibitem [{\citenamefont {Lewenstein}\ \emph {et~al.}(2007)\citenamefont
  {Lewenstein}, \citenamefont {Sanpera}, \citenamefont {Ahufinger},
  \citenamefont {Damski}, \citenamefont {Sen(De)},\ and\ \citenamefont
  {Sen}}]{Lewenstein2007}%
  \BibitemOpen
  \bibfield  {author} {\bibinfo {author} {\bibfnamefont {M.}~\bibnamefont
  {Lewenstein}}, \bibinfo {author} {\bibfnamefont {A.}~\bibnamefont {Sanpera}},
  \bibinfo {author} {\bibfnamefont {V.}~\bibnamefont {Ahufinger}}, \bibinfo
  {author} {\bibfnamefont {B.}~\bibnamefont {Damski}}, \bibinfo {author}
  {\bibfnamefont {A.}~\bibnamefont {Sen(De)}}, \ and\ \bibinfo {author}
  {\bibfnamefont {U.}~\bibnamefont {Sen}},\ }\href {\doibase
  10.1080/00018730701223200} {\bibfield  {journal} {\bibinfo  {journal}
  {Advances in Physics}\ }\textbf {\bibinfo {volume} {56}},\ \bibinfo {pages}
  {243} (\bibinfo {year} {2007})}\BibitemShut {NoStop}%
\bibitem [{\citenamefont {Ritsch}\ \emph {et~al.}(2013)\citenamefont {Ritsch},
  \citenamefont {Domokos}, \citenamefont {Brennecke},\ and\ \citenamefont
  {Esslinger}}]{Ritsch2013}%
  \BibitemOpen
  \bibfield  {author} {\bibinfo {author} {\bibfnamefont {H.}~\bibnamefont
  {Ritsch}}, \bibinfo {author} {\bibfnamefont {P.}~\bibnamefont {Domokos}},
  \bibinfo {author} {\bibfnamefont {F.}~\bibnamefont {Brennecke}}, \ and\
  \bibinfo {author} {\bibfnamefont {T.}~\bibnamefont {Esslinger}},\ }\href
  {\doibase 10.1103/RevModPhys.85.553} {\bibfield  {journal} {\bibinfo
  {journal} {Rev. Mod. Phys.}\ }\textbf {\bibinfo {volume} {85}},\ \bibinfo
  {pages} {553} (\bibinfo {year} {2013})}\BibitemShut {NoStop}%
\bibitem [{\citenamefont {Mekhov}\ and\ \citenamefont
  {Ritsch}(2012)}]{Mekhov2012}%
  \BibitemOpen
  \bibfield  {author} {\bibinfo {author} {\bibfnamefont {I.~B.}\ \bibnamefont
  {Mekhov}}\ and\ \bibinfo {author} {\bibfnamefont {H.}~\bibnamefont
  {Ritsch}},\ }\href {\doibase 10.1088/0953-4075/45/10/102001} {\bibfield
  {journal} {\bibinfo  {journal} {J. Phys. B At. Mol. Opt. Phys.}\ }\textbf
  {\bibinfo {volume} {45}},\ \bibinfo {pages} {102001} (\bibinfo {year}
  {2012})}\BibitemShut {NoStop}%
\bibitem [{\citenamefont {Ruostekoski}\ and\ \citenamefont
  {Walls}(1997)}]{Ruostekoski1997}%
  \BibitemOpen
  \bibfield  {author} {\bibinfo {author} {\bibfnamefont {J.}~\bibnamefont
  {Ruostekoski}}\ and\ \bibinfo {author} {\bibfnamefont {D.~F.}\ \bibnamefont
  {Walls}},\ }\href@noop {} {\bibfield  {journal} {\bibinfo  {journal}
  {Physical Review A}\ }\textbf {\bibinfo {volume} {56}},\ \bibinfo {pages}
  {2996} (\bibinfo {year} {1997})}\BibitemShut {NoStop}%
\bibitem [{\citenamefont {Mekhov}\ \emph
  {et~al.}(2007{\natexlab{a}})\citenamefont {Mekhov}, \citenamefont
  {Maschler},\ and\ \citenamefont {Ritsch}}]{Mekhov2007NP}%
  \BibitemOpen
  \bibfield  {author} {\bibinfo {author} {\bibfnamefont {I.~B.}\ \bibnamefont
  {Mekhov}}, \bibinfo {author} {\bibfnamefont {C.}~\bibnamefont {Maschler}}, \
  and\ \bibinfo {author} {\bibfnamefont {H.}~\bibnamefont {Ritsch}},\
  }\href@noop {} {\bibfield  {journal} {\bibinfo  {journal} {Nat. Phys.}\
  }\textbf {\bibinfo {volume} {3}},\ \bibinfo {pages} {319} (\bibinfo {year}
  {2007}{\natexlab{a}})}\BibitemShut {NoStop}%
\bibitem [{\citenamefont {Chen}\ and\ \citenamefont
  {Meystre}(2009)}]{Chen2009}%
  \BibitemOpen
  \bibfield  {author} {\bibinfo {author} {\bibfnamefont {W.}~\bibnamefont
  {Chen}}\ and\ \bibinfo {author} {\bibfnamefont {P.}~\bibnamefont {Meystre}},\
  }\href@noop {} {\bibfield  {journal} {\bibinfo  {journal} {Phys. Rev. A}\
  }\textbf {\bibinfo {volume} {79}},\ \bibinfo {pages} {043801} (\bibinfo
  {year} {2009})}\BibitemShut {NoStop}%
\bibitem [{\citenamefont {Mekhov}\ and\ \citenamefont
  {Ritsch}(2009{\natexlab{a}})}]{Mekhov2009LP}%
  \BibitemOpen
  \bibfield  {author} {\bibinfo {author} {\bibfnamefont {I.~B.}\ \bibnamefont
  {Mekhov}}\ and\ \bibinfo {author} {\bibfnamefont {H.}~\bibnamefont
  {Ritsch}},\ }\href@noop {} {\bibfield  {journal} {\bibinfo  {journal} {Laser
  physics}\ }\textbf {\bibinfo {volume} {19}},\ \bibinfo {pages} {610}
  (\bibinfo {year} {2009}{\natexlab{a}})}\BibitemShut {NoStop}%
\bibitem [{\citenamefont {Mekhov}\ and\ \citenamefont
  {Ritsch}(2010)}]{Mekhov2010LP}%
  \BibitemOpen
  \bibfield  {author} {\bibinfo {author} {\bibfnamefont {I.~B.}\ \bibnamefont
  {Mekhov}}\ and\ \bibinfo {author} {\bibfnamefont {H.}~\bibnamefont
  {Ritsch}},\ }\href@noop {} {\bibfield  {journal} {\bibinfo  {journal} {Laser
  physics}\ }\textbf {\bibinfo {volume} {20}},\ \bibinfo {pages} {694}
  (\bibinfo {year} {2010})}\BibitemShut {NoStop}%
\bibitem [{\citenamefont {Mekhov}\ and\ \citenamefont
  {Ritsch}(2011)}]{Mekhov2011LP}%
  \BibitemOpen
  \bibfield  {author} {\bibinfo {author} {\bibfnamefont {I.~B.}\ \bibnamefont
  {Mekhov}}\ and\ \bibinfo {author} {\bibfnamefont {H.}~\bibnamefont
  {Ritsch}},\ }\href@noop {} {\bibfield  {journal} {\bibinfo  {journal} {Laser
  Physics}\ }\textbf {\bibinfo {volume} {21}},\ \bibinfo {pages} {1486}
  (\bibinfo {year} {2011})}\BibitemShut {NoStop}%
\bibitem [{\citenamefont {Mekhov}(2013)}]{Mekhov2013LP}%
  \BibitemOpen
  \bibfield  {author} {\bibinfo {author} {\bibfnamefont {I.~B.}\ \bibnamefont
  {Mekhov}},\ }\href@noop {} {\bibfield  {journal} {\bibinfo  {journal} {Laser
  Physics}\ }\textbf {\bibinfo {volume} {23}},\ \bibinfo {pages} {015501}
  (\bibinfo {year} {2013})}\BibitemShut {NoStop}%
\bibitem [{\citenamefont {Pedersen}\ \emph {et~al.}(2014)\citenamefont
  {Pedersen}, \citenamefont {S{\o}rensen}, \citenamefont {Tichy},\ and\
  \citenamefont {Sherson}}]{Pedersen2014}%
  \BibitemOpen
  \bibfield  {author} {\bibinfo {author} {\bibfnamefont {M.~K.}\ \bibnamefont
  {Pedersen}}, \bibinfo {author} {\bibfnamefont {J.~J.~W.}\ \bibnamefont
  {S{\o}rensen}}, \bibinfo {author} {\bibfnamefont {M.~C.}\ \bibnamefont
  {Tichy}}, \ and\ \bibinfo {author} {\bibfnamefont {J.~F.}\ \bibnamefont
  {Sherson}},\ }\href@noop {} {\bibfield  {journal} {\bibinfo  {journal} {New
  Journal of Physics}\ }\textbf {\bibinfo {volume} {16}},\ \bibinfo {pages}
  {113038} (\bibinfo {year} {2014})}\BibitemShut {NoStop}%
\bibitem [{\citenamefont {Elliott}\ \emph
  {et~al.}(2015{\natexlab{a}})\citenamefont {Elliott}, \citenamefont
  {Kozlowski}, \citenamefont {Caballero-Benitez},\ and\ \citenamefont
  {Mekhov}}]{Elliott2015}%
  \BibitemOpen
  \bibfield  {author} {\bibinfo {author} {\bibfnamefont {T.~J.}\ \bibnamefont
  {Elliott}}, \bibinfo {author} {\bibfnamefont {W.}~\bibnamefont {Kozlowski}},
  \bibinfo {author} {\bibfnamefont {S.~F.}\ \bibnamefont {Caballero-Benitez}},
  \ and\ \bibinfo {author} {\bibfnamefont {I.~B.}\ \bibnamefont {Mekhov}},\
  }\href@noop {} {\bibfield  {journal} {\bibinfo  {journal} {Phys. Rev. Lett.}\
  }\textbf {\bibinfo {volume} {114}},\ \bibinfo {pages} {113604} (\bibinfo
  {year} {2015}{\natexlab{a}})}\BibitemShut {NoStop}%
\bibitem [{\citenamefont {Kozlowski}\ \emph {et~al.}(2015)\citenamefont
  {Kozlowski}, \citenamefont {Caballero-Benitez},\ and\ \citenamefont
  {Mekhov}}]{Kozlowski2015}%
  \BibitemOpen
  \bibfield  {author} {\bibinfo {author} {\bibfnamefont {W.}~\bibnamefont
  {Kozlowski}}, \bibinfo {author} {\bibfnamefont {S.~F.}\ \bibnamefont
  {Caballero-Benitez}}, \ and\ \bibinfo {author} {\bibfnamefont {I.~B.}\
  \bibnamefont {Mekhov}},\ }\href@noop {} {\bibfield  {journal} {\bibinfo
  {journal} {Phys. Rev. A}\ }\textbf {\bibinfo {volume} {92}},\ \bibinfo
  {pages} {013613} (\bibinfo {year} {2015})}\BibitemShut {NoStop}%
\bibitem [{\citenamefont {Mazzucchi}\ \emph
  {et~al.}(2016{\natexlab{a}})\citenamefont {Mazzucchi}, \citenamefont
  {Kozlowski}, \citenamefont {Caballero-Benitez}, \citenamefont {Elliott},\
  and\ \citenamefont {Mekhov}}]{Mazzucchi2015}%
  \BibitemOpen
  \bibfield  {author} {\bibinfo {author} {\bibfnamefont {G.}~\bibnamefont
  {Mazzucchi}}, \bibinfo {author} {\bibfnamefont {W.}~\bibnamefont
  {Kozlowski}}, \bibinfo {author} {\bibfnamefont {S.~F.}\ \bibnamefont
  {Caballero-Benitez}}, \bibinfo {author} {\bibfnamefont {T.~J.}\ \bibnamefont
  {Elliott}}, \ and\ \bibinfo {author} {\bibfnamefont {I.~B.}\ \bibnamefont
  {Mekhov}},\ }\href@noop {} {\bibfield  {journal} {\bibinfo  {journal} {Phys.
  Rev. A}\ }\textbf {\bibinfo {volume} {93}},\ \bibinfo {pages} {023632}
  (\bibinfo {year} {2016}{\natexlab{a}})}\BibitemShut {NoStop}%
\bibitem [{\citenamefont {Elliott}\ \emph
  {et~al.}(2015{\natexlab{b}})\citenamefont {Elliott}, \citenamefont
  {Mazzucchi}, \citenamefont {Kozlowski}, \citenamefont {Caballero-Benitez},\
  and\ \citenamefont {Mekhov}}]{Atom2015}%
  \BibitemOpen
  \bibfield  {author} {\bibinfo {author} {\bibfnamefont {T.~J.}\ \bibnamefont
  {Elliott}}, \bibinfo {author} {\bibfnamefont {G.}~\bibnamefont {Mazzucchi}},
  \bibinfo {author} {\bibfnamefont {W.}~\bibnamefont {Kozlowski}}, \bibinfo
  {author} {\bibfnamefont {S.~F.}\ \bibnamefont {Caballero-Benitez}}, \ and\
  \bibinfo {author} {\bibfnamefont {I.~B.}\ \bibnamefont {Mekhov}},\
  }\href@noop {} {\bibfield  {journal} {\bibinfo  {journal} {Atoms}\ }\textbf
  {\bibinfo {volume} {3}},\ \bibinfo {pages} {392} (\bibinfo {year}
  {2015}{\natexlab{b}})}\BibitemShut {NoStop}%
\bibitem [{\citenamefont {Wade}\ \emph {et~al.}(2015)\citenamefont {Wade},
  \citenamefont {Sherson},\ and\ \citenamefont {M\o{}lmer}}]{Wade2015}%
  \BibitemOpen
  \bibfield  {author} {\bibinfo {author} {\bibfnamefont {A.~C.~J.}\
  \bibnamefont {Wade}}, \bibinfo {author} {\bibfnamefont {J.~F.}\ \bibnamefont
  {Sherson}}, \ and\ \bibinfo {author} {\bibfnamefont {K.}~\bibnamefont
  {M\o{}lmer}},\ }\href@noop {} {\bibfield  {journal} {\bibinfo  {journal}
  {Phys. Rev. Lett.}\ }\textbf {\bibinfo {volume} {115}},\ \bibinfo {pages}
  {060401} (\bibinfo {year} {2015})}\BibitemShut {NoStop}%
\bibitem [{\citenamefont {Ashida}\ and\ \citenamefont
  {Ueda}(2015)}]{Ashida2015}%
  \BibitemOpen
  \bibfield  {author} {\bibinfo {author} {\bibfnamefont {Y.}~\bibnamefont
  {Ashida}}\ and\ \bibinfo {author} {\bibfnamefont {M.}~\bibnamefont {Ueda}},\
  }\href@noop {} {\bibfield  {journal} {\bibinfo  {journal} {Phys. Rev. Lett.}\
  }\textbf {\bibinfo {volume} {115}},\ \bibinfo {pages} {095301} (\bibinfo
  {year} {2015})}\BibitemShut {NoStop}%
\bibitem [{\citenamefont {Mazzucchi}\ \emph
  {et~al.}(2016{\natexlab{b}})\citenamefont {Mazzucchi}, \citenamefont
  {Kozlowski}, \citenamefont {Caballero-Benitez},\ and\ \citenamefont
  {Mekhov}}]{Mazzucchi2016}%
  \BibitemOpen
  \bibfield  {author} {\bibinfo {author} {\bibfnamefont {G.}~\bibnamefont
  {Mazzucchi}}, \bibinfo {author} {\bibfnamefont {W.}~\bibnamefont
  {Kozlowski}}, \bibinfo {author} {\bibfnamefont {S.~F.}\ \bibnamefont
  {Caballero-Benitez}}, \ and\ \bibinfo {author} {\bibfnamefont {I.~B.}\
  \bibnamefont {Mekhov}},\ }\href
  {http://stacks.iop.org/1367-2630/18/i=7/a=073017} {\bibfield  {journal}
  {\bibinfo  {journal} {New Journal of Physics}\ }\textbf {\bibinfo {volume}
  {18}},\ \bibinfo {pages} {073017} (\bibinfo {year}
  {2016}{\natexlab{b}})}\BibitemShut {NoStop}%
\bibitem [{\citenamefont {Caballero-Benitez}\ \emph {et~al.}(2016)\citenamefont
  {Caballero-Benitez}, \citenamefont {Mazzucchi},\ and\ \citenamefont
  {Mekhov}}]{Caballero2016}%
  \BibitemOpen
  \bibfield  {author} {\bibinfo {author} {\bibfnamefont {S.~F.}\ \bibnamefont
  {Caballero-Benitez}}, \bibinfo {author} {\bibfnamefont {G.}~\bibnamefont
  {Mazzucchi}}, \ and\ \bibinfo {author} {\bibfnamefont {I.~B.}\ \bibnamefont
  {Mekhov}},\ }\href {\doibase 10.1103/PhysRevA.93.063632} {\bibfield
  {journal} {\bibinfo  {journal} {Phys. Rev. A}\ }\textbf {\bibinfo {volume}
  {93}},\ \bibinfo {pages} {063632} (\bibinfo {year} {2016})}\BibitemShut
  {NoStop}%
\bibitem [{\citenamefont {Baumann}\ \emph {et~al.}(2010)\citenamefont
  {Baumann}, \citenamefont {Guerlin}, \citenamefont {Brennecke},\ and\
  \citenamefont {Esslinger}}]{Baumann2010}%
  \BibitemOpen
  \bibfield  {author} {\bibinfo {author} {\bibfnamefont {K.}~\bibnamefont
  {Baumann}}, \bibinfo {author} {\bibfnamefont {C.}~\bibnamefont {Guerlin}},
  \bibinfo {author} {\bibfnamefont {F.}~\bibnamefont {Brennecke}}, \ and\
  \bibinfo {author} {\bibfnamefont {T.}~\bibnamefont {Esslinger}},\ }\href@noop
  {} {\bibfield  {journal} {\bibinfo  {journal} {Nature}\ }\textbf {\bibinfo
  {volume} {464}},\ \bibinfo {pages} {1301} (\bibinfo {year}
  {2010})}\BibitemShut {NoStop}%
\bibitem [{\citenamefont {Wolke}\ \emph {et~al.}(2012)\citenamefont {Wolke},
  \citenamefont {Klinner}, \citenamefont {Ke{\ss}ler},\ and\ \citenamefont
  {Hemmerich}}]{Wolke2012}%
  \BibitemOpen
  \bibfield  {author} {\bibinfo {author} {\bibfnamefont {M.}~\bibnamefont
  {Wolke}}, \bibinfo {author} {\bibfnamefont {J.}~\bibnamefont {Klinner}},
  \bibinfo {author} {\bibfnamefont {H.}~\bibnamefont {Ke{\ss}ler}}, \ and\
  \bibinfo {author} {\bibfnamefont {A.}~\bibnamefont {Hemmerich}},\ }\href@noop
  {} {\bibfield  {journal} {\bibinfo  {journal} {Science}\ }\textbf {\bibinfo
  {volume} {337}},\ \bibinfo {pages} {75} (\bibinfo {year} {2012})}\BibitemShut
  {NoStop}%
\bibitem [{\citenamefont {Schmidt}\ \emph {et~al.}(2014)\citenamefont
  {Schmidt}, \citenamefont {Tomczyk}, \citenamefont {Slama},\ and\
  \citenamefont {Zimmermann}}]{Schmidt2014}%
  \BibitemOpen
  \bibfield  {author} {\bibinfo {author} {\bibfnamefont {D.}~\bibnamefont
  {Schmidt}}, \bibinfo {author} {\bibfnamefont {H.}~\bibnamefont {Tomczyk}},
  \bibinfo {author} {\bibfnamefont {S.}~\bibnamefont {Slama}}, \ and\ \bibinfo
  {author} {\bibfnamefont {C.}~\bibnamefont {Zimmermann}},\ }\href@noop {}
  {\bibfield  {journal} {\bibinfo  {journal} {Phys. Rev. Lett.}\ }\textbf
  {\bibinfo {volume} {112}},\ \bibinfo {pages} {115302} (\bibinfo {year}
  {2014})}\BibitemShut {NoStop}%
\bibitem [{\citenamefont {Miyake}\ \emph {et~al.}(2011)\citenamefont {Miyake},
  \citenamefont {Siviloglou}, \citenamefont {Puentes}, \citenamefont
  {Pritchard}, \citenamefont {Ketterle},\ and\ \citenamefont
  {Weld}}]{Miyake2011}%
  \BibitemOpen
  \bibfield  {author} {\bibinfo {author} {\bibfnamefont {H.}~\bibnamefont
  {Miyake}}, \bibinfo {author} {\bibfnamefont {G.~A.}\ \bibnamefont
  {Siviloglou}}, \bibinfo {author} {\bibfnamefont {G.}~\bibnamefont {Puentes}},
  \bibinfo {author} {\bibfnamefont {D.~E.}\ \bibnamefont {Pritchard}}, \bibinfo
  {author} {\bibfnamefont {W.}~\bibnamefont {Ketterle}}, \ and\ \bibinfo
  {author} {\bibfnamefont {D.~M.}\ \bibnamefont {Weld}},\ }\href@noop {}
  {\bibfield  {journal} {\bibinfo  {journal} {Phys. Rev. Lett.}\ }\textbf
  {\bibinfo {volume} {107}},\ \bibinfo {pages} {175302} (\bibinfo {year}
  {2011})}\BibitemShut {NoStop}%
\bibitem [{\citenamefont {Weitenberg}\ \emph {et~al.}(2011)\citenamefont
  {Weitenberg}, \citenamefont {Endres}, \citenamefont {Sherson}, \citenamefont
  {Cheneau}, \citenamefont {Schauss}, \citenamefont {Fukuhara}, \citenamefont
  {Bloch},\ and\ \citenamefont {Kuhr}}]{Weitenberg2011}%
  \BibitemOpen
  \bibfield  {author} {\bibinfo {author} {\bibfnamefont {C.}~\bibnamefont
  {Weitenberg}}, \bibinfo {author} {\bibfnamefont {M.}~\bibnamefont {Endres}},
  \bibinfo {author} {\bibfnamefont {J.~F.}\ \bibnamefont {Sherson}}, \bibinfo
  {author} {\bibfnamefont {M.}~\bibnamefont {Cheneau}}, \bibinfo {author}
  {\bibfnamefont {P.}~\bibnamefont {Schauss}}, \bibinfo {author} {\bibfnamefont
  {T.}~\bibnamefont {Fukuhara}}, \bibinfo {author} {\bibfnamefont
  {I.}~\bibnamefont {Bloch}}, \ and\ \bibinfo {author} {\bibfnamefont
  {S.}~\bibnamefont {Kuhr}},\ }\href@noop {} {\bibfield  {journal} {\bibinfo
  {journal} {Nature}\ }\textbf {\bibinfo {volume} {471}},\ \bibinfo {pages}
  {319} (\bibinfo {year} {2011})}\BibitemShut {NoStop}%
\bibitem [{\citenamefont {Landig}\ \emph {et~al.}(2016)\citenamefont {Landig},
  \citenamefont {Hruby}, \citenamefont {Dogra}, \citenamefont {Landini},
  \citenamefont {Mottl}, \citenamefont {Donner},\ and\ \citenamefont
  {Esslinger}}]{Landig2015a}%
  \BibitemOpen
  \bibfield  {author} {\bibinfo {author} {\bibfnamefont {R.}~\bibnamefont
  {Landig}}, \bibinfo {author} {\bibfnamefont {L.}~\bibnamefont {Hruby}},
  \bibinfo {author} {\bibfnamefont {N.}~\bibnamefont {Dogra}}, \bibinfo
  {author} {\bibfnamefont {M.}~\bibnamefont {Landini}}, \bibinfo {author}
  {\bibfnamefont {R.}~\bibnamefont {Mottl}}, \bibinfo {author} {\bibfnamefont
  {T.}~\bibnamefont {Donner}}, \ and\ \bibinfo {author} {\bibfnamefont
  {T.}~\bibnamefont {Esslinger}},\ }\href@noop {} {\bibfield  {journal}
  {\bibinfo  {journal} {Nature}\ } (\bibinfo {year} {2016})}\BibitemShut
  {NoStop}%
\bibitem [{\citenamefont {Klinder}\ \emph {et~al.}(2015)\citenamefont
  {Klinder}, \citenamefont {Ke{\ss}ler}, \citenamefont {Bakhtiari},
  \citenamefont {Thorwart},\ and\ \citenamefont {Hemmerich}}]{Klinder2015}%
  \BibitemOpen
  \bibfield  {author} {\bibinfo {author} {\bibfnamefont {J.}~\bibnamefont
  {Klinder}}, \bibinfo {author} {\bibfnamefont {H.}~\bibnamefont {Ke{\ss}ler}},
  \bibinfo {author} {\bibfnamefont {M.~R.}\ \bibnamefont {Bakhtiari}}, \bibinfo
  {author} {\bibfnamefont {M.}~\bibnamefont {Thorwart}}, \ and\ \bibinfo
  {author} {\bibfnamefont {A.}~\bibnamefont {Hemmerich}},\ }\href@noop {}
  {\bibfield  {journal} {\bibinfo  {journal} {Physical Review Letters}\
  }\textbf {\bibinfo {volume} {115}},\ \bibinfo {pages} {230403} (\bibinfo
  {year} {2015})}\BibitemShut {NoStop}%
\bibitem [{\citenamefont {Mekhov}\ \emph
  {et~al.}(2007{\natexlab{b}})\citenamefont {Mekhov}, \citenamefont
  {Maschler},\ and\ \citenamefont {Ritsch}}]{Mekhov2007}%
  \BibitemOpen
  \bibfield  {author} {\bibinfo {author} {\bibfnamefont {I.~B.}\ \bibnamefont
  {Mekhov}}, \bibinfo {author} {\bibfnamefont {C.}~\bibnamefont {Maschler}}, \
  and\ \bibinfo {author} {\bibfnamefont {H.}~\bibnamefont {Ritsch}},\ }\href
  {\doibase 10.1103/PhysRevLett.98.100402} {\bibfield  {journal} {\bibinfo
  {journal} {Phys. Rev. Lett.}\ }\textbf {\bibinfo {volume} {98}},\ \bibinfo
  {pages} {100402} (\bibinfo {year} {2007}{\natexlab{b}})}\BibitemShut
  {NoStop}%
\bibitem [{\citenamefont {Mekhov}\ \emph
  {et~al.}(2007{\natexlab{c}})\citenamefont {Mekhov}, \citenamefont
  {Maschler},\ and\ \citenamefont {Ritsch}}]{Mekhov2007PRA}%
  \BibitemOpen
  \bibfield  {author} {\bibinfo {author} {\bibfnamefont {I.~B.}\ \bibnamefont
  {Mekhov}}, \bibinfo {author} {\bibfnamefont {C.}~\bibnamefont {Maschler}}, \
  and\ \bibinfo {author} {\bibfnamefont {H.}~\bibnamefont {Ritsch}},\
  }\href@noop {} {\bibfield  {journal} {\bibinfo  {journal} {Phys. Rev. A}\
  }\textbf {\bibinfo {volume} {76}},\ \bibinfo {pages} {053618} (\bibinfo
  {year} {2007}{\natexlab{c}})}\BibitemShut {NoStop}%
\bibitem [{\citenamefont {Eckert}\ \emph {et~al.}(2008)\citenamefont {Eckert},
  \citenamefont {Romero-Isart}, \citenamefont {Rodriguez}, \citenamefont
  {Lewenstein}, \citenamefont {Polzik},\ and\ \citenamefont
  {Sanpera}}]{PolzikNatPh2008}%
  \BibitemOpen
  \bibfield  {author} {\bibinfo {author} {\bibfnamefont {K.}~\bibnamefont
  {Eckert}}, \bibinfo {author} {\bibfnamefont {O.}~\bibnamefont
  {Romero-Isart}}, \bibinfo {author} {\bibfnamefont {M.}~\bibnamefont
  {Rodriguez}}, \bibinfo {author} {\bibfnamefont {M.}~\bibnamefont
  {Lewenstein}}, \bibinfo {author} {\bibfnamefont {E.~S.}\ \bibnamefont
  {Polzik}}, \ and\ \bibinfo {author} {\bibfnamefont {A.}~\bibnamefont
  {Sanpera}},\ }\href@noop {} {\bibfield  {journal} {\bibinfo  {journal} {Nat.
  Phys.}\ }\textbf {\bibinfo {volume} {4}},\ \bibinfo {pages} {50} (\bibinfo
  {year} {2008})}\BibitemShut {NoStop}%
\bibitem [{\citenamefont {Roscilde}\ \emph {et~al.}(2009)\citenamefont
  {Roscilde}, \citenamefont {Rodríguez}, \citenamefont {Eckert}, \citenamefont
  {Romero-Isart}, \citenamefont {Lewenstein}, \citenamefont {Polzik},\ and\
  \citenamefont {Sanpera}}]{Roscilde2009}%
  \BibitemOpen
  \bibfield  {author} {\bibinfo {author} {\bibfnamefont {T.}~\bibnamefont
  {Roscilde}}, \bibinfo {author} {\bibfnamefont {M.}~\bibnamefont
  {Rodríguez}}, \bibinfo {author} {\bibfnamefont {K.}~\bibnamefont {Eckert}},
  \bibinfo {author} {\bibfnamefont {O.}~\bibnamefont {Romero-Isart}}, \bibinfo
  {author} {\bibfnamefont {M.}~\bibnamefont {Lewenstein}}, \bibinfo {author}
  {\bibfnamefont {E.}~\bibnamefont {Polzik}}, \ and\ \bibinfo {author}
  {\bibfnamefont {A.}~\bibnamefont {Sanpera}},\ }\href@noop {} {\bibfield
  {journal} {\bibinfo  {journal} {New Jour. Phys.}\ }\textbf {\bibinfo {volume}
  {11}},\ \bibinfo {pages} {055041} (\bibinfo {year} {2009})}\BibitemShut
  {NoStop}%
\bibitem [{\citenamefont {Mekhov}\ and\ \citenamefont
  {Ritsch}(2009{\natexlab{b}})}]{Mekhov2009}%
  \BibitemOpen
  \bibfield  {author} {\bibinfo {author} {\bibfnamefont {I.~B.}\ \bibnamefont
  {Mekhov}}\ and\ \bibinfo {author} {\bibfnamefont {H.}~\bibnamefont
  {Ritsch}},\ }\href {\doibase 10.1103/PhysRevLett.102.020403} {\bibfield
  {journal} {\bibinfo  {journal} {Phys. Rev. Lett.}\ }\textbf {\bibinfo
  {volume} {102}},\ \bibinfo {pages} {020403} (\bibinfo {year}
  {2009}{\natexlab{b}})}\BibitemShut {NoStop}%
\bibitem [{\citenamefont {Mekhov}\ and\ \citenamefont
  {Ritsch}(2009{\natexlab{c}})}]{Mekhov2009PRA}%
  \BibitemOpen
  \bibfield  {author} {\bibinfo {author} {\bibfnamefont {I.~B.}\ \bibnamefont
  {Mekhov}}\ and\ \bibinfo {author} {\bibfnamefont {H.}~\bibnamefont
  {Ritsch}},\ }\href@noop {} {\bibfield  {journal} {\bibinfo  {journal} {Phys.
  Rev. A}\ }\textbf {\bibinfo {volume} {80}},\ \bibinfo {pages} {013604}
  (\bibinfo {year} {2009}{\natexlab{c}})}\BibitemShut {NoStop}%
\bibitem [{\citenamefont {De~Chiara}\ \emph {et~al.}(2011)\citenamefont
  {De~Chiara}, \citenamefont {Romero-Isart},\ and\ \citenamefont
  {Sanpera}}]{DeChiara2011}%
  \BibitemOpen
  \bibfield  {author} {\bibinfo {author} {\bibfnamefont {G.}~\bibnamefont
  {De~Chiara}}, \bibinfo {author} {\bibfnamefont {O.}~\bibnamefont
  {Romero-Isart}}, \ and\ \bibinfo {author} {\bibfnamefont {A.}~\bibnamefont
  {Sanpera}},\ }\href@noop {} {\bibfield  {journal} {\bibinfo  {journal} {Phys.
  Rev. A}\ }\textbf {\bibinfo {volume} {83}},\ \bibinfo {pages} {021604}
  (\bibinfo {year} {2011})}\BibitemShut {NoStop}%
\bibitem [{\citenamefont {Hauke}\ \emph {et~al.}(2013)\citenamefont {Hauke},
  \citenamefont {Sewell}, \citenamefont {Mitchell},\ and\ \citenamefont
  {Lewenstein}}]{Hauke2013}%
  \BibitemOpen
  \bibfield  {author} {\bibinfo {author} {\bibfnamefont {P.}~\bibnamefont
  {Hauke}}, \bibinfo {author} {\bibfnamefont {R.~J.}\ \bibnamefont {Sewell}},
  \bibinfo {author} {\bibfnamefont {M.~W.}\ \bibnamefont {Mitchell}}, \ and\
  \bibinfo {author} {\bibfnamefont {M.}~\bibnamefont {Lewenstein}},\
  }\href@noop {} {\bibfield  {journal} {\bibinfo  {journal} {Phys. Rev. A}\
  }\textbf {\bibinfo {volume} {87}},\ \bibinfo {pages} {021601} (\bibinfo
  {year} {2013})}\BibitemShut {NoStop}%
\bibitem [{\citenamefont {Rogers}\ \emph {et~al.}(2014)\citenamefont {Rogers},
  \citenamefont {Paternostro}, \citenamefont {Sherson},\ and\ \citenamefont
  {De~Chiara}}]{Rogers2014}%
  \BibitemOpen
  \bibfield  {author} {\bibinfo {author} {\bibfnamefont {B.}~\bibnamefont
  {Rogers}}, \bibinfo {author} {\bibfnamefont {M.}~\bibnamefont {Paternostro}},
  \bibinfo {author} {\bibfnamefont {J.~F.}\ \bibnamefont {Sherson}}, \ and\
  \bibinfo {author} {\bibfnamefont {G.}~\bibnamefont {De~Chiara}},\ }\href
  {\doibase 10.1103/PhysRevA.90.043618} {\bibfield  {journal} {\bibinfo
  {journal} {Phys. Rev. A}\ }\textbf {\bibinfo {volume} {90}},\ \bibinfo
  {pages} {043618} (\bibinfo {year} {2014})}\BibitemShut {NoStop}%
\bibitem [{\citenamefont {Nagy}\ \emph {et~al.}(2008)\citenamefont {Nagy},
  \citenamefont {Szirmai},\ and\ \citenamefont {Domokos}}]{Nagy2008}%
  \BibitemOpen
  \bibfield  {author} {\bibinfo {author} {\bibfnamefont {D.}~\bibnamefont
  {Nagy}}, \bibinfo {author} {\bibfnamefont {G.}~\bibnamefont {Szirmai}}, \
  and\ \bibinfo {author} {\bibfnamefont {P.}~\bibnamefont {Domokos}},\
  }\href@noop {} {\bibfield  {journal} {\bibinfo  {journal} {The European
  Physical Journal D}\ }\textbf {\bibinfo {volume} {48}},\ \bibinfo {pages}
  {127} (\bibinfo {year} {2008})}\BibitemShut {NoStop}%
\bibitem [{\citenamefont {Niedenzu}\ \emph {et~al.}(2013)\citenamefont
  {Niedenzu}, \citenamefont {Sch{\"u}tz}, \citenamefont {Habibian},
  \citenamefont {Morigi},\ and\ \citenamefont {Ritsch}}]{Niedenzu2013}%
  \BibitemOpen
  \bibfield  {author} {\bibinfo {author} {\bibfnamefont {W.}~\bibnamefont
  {Niedenzu}}, \bibinfo {author} {\bibfnamefont {S.}~\bibnamefont
  {Sch{\"u}tz}}, \bibinfo {author} {\bibfnamefont {H.}~\bibnamefont
  {Habibian}}, \bibinfo {author} {\bibfnamefont {G.}~\bibnamefont {Morigi}}, \
  and\ \bibinfo {author} {\bibfnamefont {H.}~\bibnamefont {Ritsch}},\
  }\href@noop {} {\bibfield  {journal} {\bibinfo  {journal} {Physical Review
  A}\ }\textbf {\bibinfo {volume} {88}},\ \bibinfo {pages} {033830} (\bibinfo
  {year} {2013})}\BibitemShut {NoStop}%
\bibitem [{\citenamefont {Bakhtiari}\ \emph {et~al.}(2015)\citenamefont
  {Bakhtiari}, \citenamefont {Hemmerich}, \citenamefont {Ritsch},\ and\
  \citenamefont {Thorwart}}]{Bakhtiari2015}%
  \BibitemOpen
  \bibfield  {author} {\bibinfo {author} {\bibfnamefont {M.~R.}\ \bibnamefont
  {Bakhtiari}}, \bibinfo {author} {\bibfnamefont {A.}~\bibnamefont
  {Hemmerich}}, \bibinfo {author} {\bibfnamefont {H.}~\bibnamefont {Ritsch}}, \
  and\ \bibinfo {author} {\bibfnamefont {M.}~\bibnamefont {Thorwart}},\
  }\href@noop {} {\bibfield  {journal} {\bibinfo  {journal} {Phys. Rev. Lett.}\
  }\textbf {\bibinfo {volume} {114}},\ \bibinfo {pages} {123601} (\bibinfo
  {year} {2015})}\BibitemShut {NoStop}%
\bibitem [{\citenamefont {Ivanov}\ and\ \citenamefont
  {Ivanova}(2015)}]{Ivanov2015}%
  \BibitemOpen
  \bibfield  {author} {\bibinfo {author} {\bibfnamefont {D.}~\bibnamefont
  {Ivanov}}\ and\ \bibinfo {author} {\bibfnamefont {T.~Y.}\ \bibnamefont
  {Ivanova}},\ }\href@noop {} {\bibfield  {journal} {\bibinfo  {journal} {J.
  Exp. Theor. Phys.}\ }\textbf {\bibinfo {volume} {121}},\ \bibinfo {pages}
  {179} (\bibinfo {year} {2015})}\BibitemShut {NoStop}%
\bibitem [{\citenamefont {Sch{\"u}tz}\ \emph {et~al.}(2015)\citenamefont
  {Sch{\"u}tz}, \citenamefont {J{\"a}ger},\ and\ \citenamefont
  {Morigi}}]{Schutz2015}%
  \BibitemOpen
  \bibfield  {author} {\bibinfo {author} {\bibfnamefont {S.}~\bibnamefont
  {Sch{\"u}tz}}, \bibinfo {author} {\bibfnamefont {S.~B.}\ \bibnamefont
  {J{\"a}ger}}, \ and\ \bibinfo {author} {\bibfnamefont {G.}~\bibnamefont
  {Morigi}},\ }\href@noop {} {\bibfield  {journal} {\bibinfo  {journal}
  {arXiv:1508.06606}\ } (\bibinfo {year} {2015})}\BibitemShut {NoStop}%
\bibitem [{\citenamefont {Winterauer}\ \emph {et~al.}(2015)\citenamefont
  {Winterauer}, \citenamefont {Niedenzu},\ and\ \citenamefont
  {Ritsch}}]{Winterauer2015}%
  \BibitemOpen
  \bibfield  {author} {\bibinfo {author} {\bibfnamefont {D.~J.}\ \bibnamefont
  {Winterauer}}, \bibinfo {author} {\bibfnamefont {W.}~\bibnamefont
  {Niedenzu}}, \ and\ \bibinfo {author} {\bibfnamefont {H.}~\bibnamefont
  {Ritsch}},\ }\href@noop {} {\bibfield  {journal} {\bibinfo  {journal}
  {Physical Review A}\ }\textbf {\bibinfo {volume} {91}},\ \bibinfo {pages}
  {053829} (\bibinfo {year} {2015})}\BibitemShut {NoStop}%
\bibitem [{\citenamefont {Piazza}\ and\ \citenamefont
  {Ritsch}(2015)}]{Piazza2015}%
  \BibitemOpen
  \bibfield  {author} {\bibinfo {author} {\bibfnamefont {F.}~\bibnamefont
  {Piazza}}\ and\ \bibinfo {author} {\bibfnamefont {H.}~\bibnamefont
  {Ritsch}},\ }\href@noop {} {\bibfield  {journal} {\bibinfo  {journal} {Phys.
  Rev. Lett.}\ }\textbf {\bibinfo {volume} {115}},\ \bibinfo {pages} {163601}
  (\bibinfo {year} {2015})}\BibitemShut {NoStop}%
\bibitem [{\citenamefont {Wiseman}\ and\ \citenamefont
  {Milburn}(2010)}]{MeasurementControl}%
  \BibitemOpen
  \bibfield  {author} {\bibinfo {author} {\bibfnamefont {H.~M.}\ \bibnamefont
  {Wiseman}}\ and\ \bibinfo {author} {\bibfnamefont {G.~J.}\ \bibnamefont
  {Milburn}},\ }\href@noop {} {\emph {\bibinfo {title} {{Quantum Measurement
  and Control}}}}\ (\bibinfo  {publisher} {Cambridge University Press},\
  \bibinfo {year} {2010})\BibitemShut {NoStop}%
\bibitem [{\citenamefont {Diehl}\ \emph {et~al.}(2008)\citenamefont {Diehl},
  \citenamefont {Micheli}, \citenamefont {Kantian}, \citenamefont {Kraus},
  \citenamefont {B{\"u}chler},\ and\ \citenamefont {Zoller}}]{Diehl2008}%
  \BibitemOpen
  \bibfield  {author} {\bibinfo {author} {\bibfnamefont {S.}~\bibnamefont
  {Diehl}}, \bibinfo {author} {\bibfnamefont {A.}~\bibnamefont {Micheli}},
  \bibinfo {author} {\bibfnamefont {A.}~\bibnamefont {Kantian}}, \bibinfo
  {author} {\bibfnamefont {B.}~\bibnamefont {Kraus}}, \bibinfo {author}
  {\bibfnamefont {H.}~\bibnamefont {B{\"u}chler}}, \ and\ \bibinfo {author}
  {\bibfnamefont {P.}~\bibnamefont {Zoller}},\ }\href@noop {} {\bibfield
  {journal} {\bibinfo  {journal} {Nat. Phys.}\ }\textbf {\bibinfo {volume}
  {4}},\ \bibinfo {pages} {878} (\bibinfo {year} {2008})}\BibitemShut {NoStop}%
\bibitem [{\citenamefont {Bux}\ \emph {et~al.}(2013)\citenamefont {Bux},
  \citenamefont {Tomczyk}, \citenamefont {Schmidt}, \citenamefont {Courteille},
  \citenamefont {Piovella},\ and\ \citenamefont {Zimmermann}}]{Bux2013}%
  \BibitemOpen
  \bibfield  {author} {\bibinfo {author} {\bibfnamefont {S.}~\bibnamefont
  {Bux}}, \bibinfo {author} {\bibfnamefont {H.}~\bibnamefont {Tomczyk}},
  \bibinfo {author} {\bibfnamefont {D.}~\bibnamefont {Schmidt}}, \bibinfo
  {author} {\bibfnamefont {P.~W.}\ \bibnamefont {Courteille}}, \bibinfo
  {author} {\bibfnamefont {N.}~\bibnamefont {Piovella}}, \ and\ \bibinfo
  {author} {\bibfnamefont {C.}~\bibnamefont {Zimmermann}},\ }\href@noop {}
  {\bibfield  {journal} {\bibinfo  {journal} {Phys. Rev. A}\ }\textbf {\bibinfo
  {volume} {87}},\ \bibinfo {pages} {023607} (\bibinfo {year}
  {2013})}\BibitemShut {NoStop}%
\bibitem [{\citenamefont {Ke{\ss}ler}\ \emph {et~al.}(2014)\citenamefont
  {Ke{\ss}ler}, \citenamefont {Klinder}, \citenamefont {Wolke},\ and\
  \citenamefont {Hemmerich}}]{Kessler2014}%
  \BibitemOpen
  \bibfield  {author} {\bibinfo {author} {\bibfnamefont {H.}~\bibnamefont
  {Ke{\ss}ler}}, \bibinfo {author} {\bibfnamefont {J.}~\bibnamefont {Klinder}},
  \bibinfo {author} {\bibfnamefont {M.}~\bibnamefont {Wolke}}, \ and\ \bibinfo
  {author} {\bibfnamefont {A.}~\bibnamefont {Hemmerich}},\ }\href@noop {}
  {\bibfield  {journal} {\bibinfo  {journal} {Phys. Rev. Lett.}\ }\textbf
  {\bibinfo {volume} {113}},\ \bibinfo {pages} {070404} (\bibinfo {year}
  {2014})}\BibitemShut {NoStop}%
\bibitem [{\citenamefont {Landig}\ \emph {et~al.}(2015)\citenamefont {Landig},
  \citenamefont {Brennecke}, \citenamefont {Mottl}, \citenamefont {Donner},\
  and\ \citenamefont {Esslinger}}]{Landig2015}%
  \BibitemOpen
  \bibfield  {author} {\bibinfo {author} {\bibfnamefont {R.}~\bibnamefont
  {Landig}}, \bibinfo {author} {\bibfnamefont {F.}~\bibnamefont {Brennecke}},
  \bibinfo {author} {\bibfnamefont {R.}~\bibnamefont {Mottl}}, \bibinfo
  {author} {\bibfnamefont {T.}~\bibnamefont {Donner}}, \ and\ \bibinfo {author}
  {\bibfnamefont {T.}~\bibnamefont {Esslinger}},\ }\href@noop {} {\bibfield
  {journal} {\bibinfo  {journal} {Nat. Comms.}\ }\textbf {\bibinfo {volume}
  {6}} (\bibinfo {year} {2015})}\BibitemShut {NoStop}%
\bibitem [{\citenamefont {Caballero-Benitez}\ and\ \citenamefont
  {Mekhov}(2015{\natexlab{a}})}]{Caballero2015}%
  \BibitemOpen
  \bibfield  {author} {\bibinfo {author} {\bibfnamefont {S.~F.}\ \bibnamefont
  {Caballero-Benitez}}\ and\ \bibinfo {author} {\bibfnamefont {I.~B.}\
  \bibnamefont {Mekhov}},\ }\href {\doibase 10.1103/PhysRevLett.115.243604}
  {\bibfield  {journal} {\bibinfo  {journal} {Phys. Rev. Lett.}\ }\textbf
  {\bibinfo {volume} {115}},\ \bibinfo {pages} {243604} (\bibinfo {year}
  {2015}{\natexlab{a}})}\BibitemShut {NoStop}%
\bibitem [{\citenamefont {Caballero-Benitez}\ and\ \citenamefont
  {Mekhov}(2015{\natexlab{b}})}]{Caballero2015a}%
  \BibitemOpen
  \bibfield  {author} {\bibinfo {author} {\bibfnamefont {S.~F.}\ \bibnamefont
  {Caballero-Benitez}}\ and\ \bibinfo {author} {\bibfnamefont {I.~B.}\
  \bibnamefont {Mekhov}},\ }\href@noop {} {\bibfield  {journal} {\bibinfo
  {journal} {New Journal of Physics}\ }\textbf {\bibinfo {volume} {17}},\
  \bibinfo {pages} {123023} (\bibinfo {year} {2015}{\natexlab{b}})}\BibitemShut
  {NoStop}%
\bibitem [{\citenamefont {Pichler}\ \emph {et~al.}(2010)\citenamefont
  {Pichler}, \citenamefont {Daley},\ and\ \citenamefont
  {Zoller}}]{Pichler2010}%
  \BibitemOpen
  \bibfield  {author} {\bibinfo {author} {\bibfnamefont {H.}~\bibnamefont
  {Pichler}}, \bibinfo {author} {\bibfnamefont {A.~J.}\ \bibnamefont {Daley}},
  \ and\ \bibinfo {author} {\bibfnamefont {P.}~\bibnamefont {Zoller}},\
  }\href@noop {} {\bibfield  {journal} {\bibinfo  {journal} {Phys. Rev. A}\
  }\textbf {\bibinfo {volume} {82}},\ \bibinfo {pages} {063605} (\bibinfo
  {year} {2010})}\BibitemShut {NoStop}%
\bibitem [{\citenamefont {Sarkar}\ \emph {et~al.}(2014)\citenamefont {Sarkar},
  \citenamefont {Langer}, \citenamefont {Schachenmayer},\ and\ \citenamefont
  {Daley}}]{Sarkar2014}%
  \BibitemOpen
  \bibfield  {author} {\bibinfo {author} {\bibfnamefont {S.}~\bibnamefont
  {Sarkar}}, \bibinfo {author} {\bibfnamefont {S.}~\bibnamefont {Langer}},
  \bibinfo {author} {\bibfnamefont {J.}~\bibnamefont {Schachenmayer}}, \ and\
  \bibinfo {author} {\bibfnamefont {A.~J.}\ \bibnamefont {Daley}},\ }\href@noop
  {} {\bibfield  {journal} {\bibinfo  {journal} {Phys. Rev. A}\ }\textbf
  {\bibinfo {volume} {90}},\ \bibinfo {pages} {023618} (\bibinfo {year}
  {2014})}\BibitemShut {NoStop}%
\bibitem [{\citenamefont {Syassen}\ \emph {et~al.}(2008)\citenamefont
  {Syassen}, \citenamefont {Bauer}, \citenamefont {Lettner}, \citenamefont
  {Volz}, \citenamefont {Dietze}, \citenamefont {Garcia-Ripoll}, \citenamefont
  {Cirac}, \citenamefont {Rempe},\ and\ \citenamefont
  {D{\"u}rr}}]{Syassen2008}%
  \BibitemOpen
  \bibfield  {author} {\bibinfo {author} {\bibfnamefont {N.}~\bibnamefont
  {Syassen}}, \bibinfo {author} {\bibfnamefont {D.~M.}\ \bibnamefont {Bauer}},
  \bibinfo {author} {\bibfnamefont {M.}~\bibnamefont {Lettner}}, \bibinfo
  {author} {\bibfnamefont {T.}~\bibnamefont {Volz}}, \bibinfo {author}
  {\bibfnamefont {D.}~\bibnamefont {Dietze}}, \bibinfo {author} {\bibfnamefont
  {J.~J.}\ \bibnamefont {Garcia-Ripoll}}, \bibinfo {author} {\bibfnamefont
  {J.~I.}\ \bibnamefont {Cirac}}, \bibinfo {author} {\bibfnamefont
  {G.}~\bibnamefont {Rempe}}, \ and\ \bibinfo {author} {\bibfnamefont
  {S.}~\bibnamefont {D{\"u}rr}},\ }\href@noop {} {\bibfield  {journal}
  {\bibinfo  {journal} {Science}\ }\textbf {\bibinfo {volume} {320}},\ \bibinfo
  {pages} {1329} (\bibinfo {year} {2008})}\BibitemShut {NoStop}%
\bibitem [{\citenamefont {Kepesidis}\ and\ \citenamefont
  {Hartmann}(2012)}]{Kepesidis2012}%
  \BibitemOpen
  \bibfield  {author} {\bibinfo {author} {\bibfnamefont {K.~V.}\ \bibnamefont
  {Kepesidis}}\ and\ \bibinfo {author} {\bibfnamefont {M.~J.}\ \bibnamefont
  {Hartmann}},\ }\href@noop {} {\bibfield  {journal} {\bibinfo  {journal}
  {Phys. Rev. A}\ }\textbf {\bibinfo {volume} {85}},\ \bibinfo {pages} {063620}
  (\bibinfo {year} {2012})}\BibitemShut {NoStop}%
\bibitem [{\citenamefont {Vidanovi{\'c}}\ \emph {et~al.}(2014)\citenamefont
  {Vidanovi{\'c}}, \citenamefont {Cocks},\ and\ \citenamefont
  {Hofstetter}}]{Vidanovic2014}%
  \BibitemOpen
  \bibfield  {author} {\bibinfo {author} {\bibfnamefont {I.}~\bibnamefont
  {Vidanovi{\'c}}}, \bibinfo {author} {\bibfnamefont {D.}~\bibnamefont
  {Cocks}}, \ and\ \bibinfo {author} {\bibfnamefont {W.}~\bibnamefont
  {Hofstetter}},\ }\href@noop {} {\bibfield  {journal} {\bibinfo  {journal}
  {Phys. Rev. A}\ }\textbf {\bibinfo {volume} {89}},\ \bibinfo {pages} {053614}
  (\bibinfo {year} {2014})}\BibitemShut {NoStop}%
\bibitem [{\citenamefont {Bernier}\ \emph {et~al.}(2014)\citenamefont
  {Bernier}, \citenamefont {Poletti},\ and\ \citenamefont
  {Kollath}}]{Bernier2014}%
  \BibitemOpen
  \bibfield  {author} {\bibinfo {author} {\bibfnamefont {J.-S.}\ \bibnamefont
  {Bernier}}, \bibinfo {author} {\bibfnamefont {D.}~\bibnamefont {Poletti}}, \
  and\ \bibinfo {author} {\bibfnamefont {C.}~\bibnamefont {Kollath}},\
  }\href@noop {} {\bibfield  {journal} {\bibinfo  {journal} {Phys. Rev. B}\
  }\textbf {\bibinfo {volume} {90}},\ \bibinfo {pages} {205125} (\bibinfo
  {year} {2014})}\BibitemShut {NoStop}%
\bibitem [{\citenamefont {Daley}(2014)}]{Daley2014}%
  \BibitemOpen
  \bibfield  {author} {\bibinfo {author} {\bibfnamefont {A.~J.}\ \bibnamefont
  {Daley}},\ }\href@noop {} {\bibfield  {journal} {\bibinfo  {journal} {Adv.
  Phys.}\ }\textbf {\bibinfo {volume} {63}},\ \bibinfo {pages} {77} (\bibinfo
  {year} {2014})}\BibitemShut {NoStop}%
\bibitem [{\citenamefont {Ates}\ \emph {et~al.}(2012)\citenamefont {Ates},
  \citenamefont {Olmos}, \citenamefont {Li},\ and\ \citenamefont
  {Lesanovsky}}]{Ates2012}%
  \BibitemOpen
  \bibfield  {author} {\bibinfo {author} {\bibfnamefont {C.}~\bibnamefont
  {Ates}}, \bibinfo {author} {\bibfnamefont {B.}~\bibnamefont {Olmos}},
  \bibinfo {author} {\bibfnamefont {W.}~\bibnamefont {Li}}, \ and\ \bibinfo
  {author} {\bibfnamefont {I.}~\bibnamefont {Lesanovsky}},\ }\href@noop {}
  {\bibfield  {journal} {\bibinfo  {journal} {Phys. Rev. Lett.}\ }\textbf
  {\bibinfo {volume} {109}},\ \bibinfo {pages} {233003} (\bibinfo {year}
  {2012})}\BibitemShut {NoStop}%
\bibitem [{\citenamefont {Everest}\ \emph {et~al.}(2014)\citenamefont
  {Everest}, \citenamefont {Hush},\ and\ \citenamefont
  {Lesanovsky}}]{Everest2014}%
  \BibitemOpen
  \bibfield  {author} {\bibinfo {author} {\bibfnamefont {B.}~\bibnamefont
  {Everest}}, \bibinfo {author} {\bibfnamefont {M.~R.}\ \bibnamefont {Hush}}, \
  and\ \bibinfo {author} {\bibfnamefont {I.}~\bibnamefont {Lesanovsky}},\
  }\href@noop {} {\bibfield  {journal} {\bibinfo  {journal} {Phys. Rev. B}\
  }\textbf {\bibinfo {volume} {90}},\ \bibinfo {pages} {134306} (\bibinfo
  {year} {2014})}\BibitemShut {NoStop}%
\bibitem [{\citenamefont {Bretheau}\ \emph {et~al.}(2015)\citenamefont
  {Bretheau}, \citenamefont {Campagne-Ibarcq}, \citenamefont {Flurin},
  \citenamefont {Mallet},\ and\ \citenamefont {Huard}}]{Bretheau2015}%
  \BibitemOpen
  \bibfield  {author} {\bibinfo {author} {\bibfnamefont {L.}~\bibnamefont
  {Bretheau}}, \bibinfo {author} {\bibfnamefont {P.}~\bibnamefont
  {Campagne-Ibarcq}}, \bibinfo {author} {\bibfnamefont {E.}~\bibnamefont
  {Flurin}}, \bibinfo {author} {\bibfnamefont {F.}~\bibnamefont {Mallet}}, \
  and\ \bibinfo {author} {\bibfnamefont {B.}~\bibnamefont {Huard}},\
  }\href@noop {} {\bibfield  {journal} {\bibinfo  {journal} {Science}\ }\textbf
  {\bibinfo {volume} {348}},\ \bibinfo {pages} {776} (\bibinfo {year}
  {2015})}\BibitemShut {NoStop}%
\bibitem [{\citenamefont {Ivanov}\ and\ \citenamefont
  {Ivanova}(2014)}]{Ivanov2014}%
  \BibitemOpen
  \bibfield  {author} {\bibinfo {author} {\bibfnamefont {D.}~\bibnamefont
  {Ivanov}}\ and\ \bibinfo {author} {\bibfnamefont {T.~Y.}\ \bibnamefont
  {Ivanova}},\ }\href@noop {} {\bibfield  {journal} {\bibinfo  {journal} {JETP
  letters}\ }\textbf {\bibinfo {volume} {100}},\ \bibinfo {pages} {481}
  (\bibinfo {year} {2014})}\BibitemShut {NoStop}%
\end{thebibliography}%

\newpage

\end{document}